\documentclass[useAMS,usenatbib]{mn2e}
\usepackage{graphicx}
\usepackage{rotating}
\usepackage{amssymb}

\title{Comparison of progenitor mass estimates for the type IIP SN 2012A  }
\author[L. Tomasella et al.]
  {L.~Tomasella,$^1$\thanks{This paper is based on European Southern Observatory (ESO) NTT long-term programme, in the framework of a large international collaboration for SN research led by S.~Benetti. For the composition of the Collaboration and its scientific goals, we refer the reader to our web pages (http://graspa.oapd.inaf.it). Observing proposals: ESO 184.D.1140; TNG AOT25\_TAC49; AOT26\_TAC46; AOT27\_TAC26.  Based on observations made with the 2.56m Nordic Optical Telescope (La Palma, Spain); Copernico 1.82m Telescope (Mt.~Ekar, Asiago, Italy); 2.5m Isaac Newton Telescope (La~Palma, Spain); Observatoire de Haute-Provence 1.93m Telescope (France);  Galileo 1.22m Telescope (Pennar, Asiago, Italy); Calar Alto Observatory 2.2m Telescope (Andalucia, Spain); 3.6m ESO NTT (La Silla, Chile); 4.2m William Herschel Telescope (La~Palma, Spain); 3.6m Telescopio Nazionale Galileo (La Palma, Spain); Schmidt 67/92 Telescope (Mt.~Ekar, Asiago, Italy); CTIO Prompt Telescopes (Cerro Tololo, Chile); ESO Trappist Telescope (La~Silla, Chile), CNTAC proposals CN2011B-092, CN2012A-103, CN2012B-FT-2 and CN2013A-153; ESO REM (La~Silla, Chile). } {E.~Cappellaro,$^1$} M.~Fraser,$^2$ M.L.~Pumo,$^1$ A.~Pastorello,$^1$   \newauthor G.~Pignata,$^3$ S.~Benetti,$^1$ F.~Bufano,$^3$ M.~Dennefeld,$^4$ A.~Harutyunyan,$^5$ T.~Iijima,$^1$ \newauthor A.~Jerkstrand,$^2$ E.~Kankare,$^6$ R.~Kotak,$^2$ L.~Magill,$^2$ V.~Nascimbeni,$^7$  \newauthor  P.~Ochner,$^1$ A.~Siviero,$^7$ S.~Smartt,$^2$ J.~Sollerman,$^8$ V.~Stanishev,$^9$ F.~Taddia,$^8$ \newauthor S.~Taubenberger,$^{10}$ M.~Turatto,$^1$ S.~Valenti,$^{11}$ D.E.~Wright,$^2$ L.~Zampieri$^1$\\
  $^1$INAF, Osservatorio Astronomico di Padova, 35122 Padova, Italy\\
  $^2$Astrophysics Research Centre, School of Mathematics and Physics, Queen University Belfast, Belfast BT7 1NN, UK\\
  $^3$Universidad Andr\'es Bello, Departamento Ciencias Fisicas, Santiago de Chile, Chile\\
  $^4$Institut d'Astrophysique de Paris (IAP) and University Pierre et Marie Curie (Paris 6), Boulevard Arago F-75014 Paris, France\\
  $^5$Fundaci\'on Galileo Galilei, INAF Telescopio Nazionale Galileo, Rambla Jos\'e Ana Fern\'andez P\'erez 7, 38712 Bre\~na Baja, TF, Spain\\
  $^6$Tuorla Observatory, Department of Physics and Astronomy, University of Turku, V\"ais\"al\"antie 20, FI-21500, Piikki\"o, Finland\\
  $^7$Dipartimento di Fisica e Astronomia Galileo Galilei, Universit\'a di Padova, vicolo dell'Osservatorio 3, 35122 Padova, Italy\\
  $^8$The Oskar Klein Centre, Department of Astronomy, AlbaNova, SE-106 91 Stockholm, Sweden\\
  $^9$CENTRA Centro Multidisciplinar de Astrof\'isica, Instituto Superior T\'ecnico, Av. Rovisco Pais 1, 1049-001 Lisbon, Portugal\\
  $^{10}$Max-Planck-Institut f\"ur Astrophysik, Karl-Schwarzchild-Str. 1, 85741 Garching, Germany\\
  $^{11}$Department of Physics, University of California, Santa Barbara, Broida Hall, Mail Code 9530, Santa Barbara, CA 93106-9530, USA}

\date{Released 2013 Xxxxx XX}

\pagerange{\pageref{firstpage}--\pageref{lastpage}} \pubyear{2013}

\begin{document}

\label{firstpage}

\maketitle

\begin{abstract}

We present the one-year long observing campaign of SN~2012A which exploded in the nearby (9.8 Mpc) irregular galaxy NGC~3239. The photometric evolution is that of a normal type IIP supernova, but the plateau is shorter and the luminosity not as constant as in other supernovae of this type.
The absolute maximum magnitude, with $M_B = -16.23 \pm 0.16$ mag, is close to the average for SN~IIP. Thanks also to the strong $UV$ flux in the early phase, SN~2012A reached a peak luminosity of about 2 $\times$ 10$^{42}$ erg s$^{-1}$, which is brighter than those of other SNe with a similar $^{56}$Ni mass. The latter was estimated from the luminosity in the exponential tail of the light curve and found to be M$(^{56}{\rm Ni}) = 0.011 \pm 0.004$~M$_\odot$, which is intermediate between standard and faint SN~IIP.

The spectral evolution of SN~2012A is also typical of SN~IIP, from the early spectra dominated by a blue continuum and very broad ($\sim$~10$^{4}$~km~s$^{-1}$) Balmer lines, to the late-photospheric spectra characterized by prominent P-Cygni features of metal lines (Fe~II, Sc~II, Ba~II, Ti~II, Ca~II, Na~I~D). The photospheric  velocity is moderately low, $\sim 3 \times 10^3$ km s$^{-1}$ at 50 days, for the low optical depth metal lines.  The nebular spectrum obtained 394 days after the shock breakout shows the typical features of SNe~IIP and the strength of the [O~I] doublet suggests a progenitor of intermediate mass, similar to SN~2004et ($\sim 15\,{\rm M_{\odot}}$).

A  candidate progenitor for SN 2012A has been identified in deep, pre-explosion $K'$-band Gemini North (NIRI) images, and found to be consistent with a star with a bolometric magnitude $-7.08 \pm 0.36$ (log $L/L_{\odot} = 4.73 \pm 0.14$ dex). The magnitude of the recovered progenitor in archival images points toward a moderate-mass $10.5_{-2}^{+4.5}\,{\rm M}_\odot $ star as the precursor of SN 2012A.

The explosion parameters and progenitor mass were also estimated by means of a hydrodynamical model, fitting the bolometric light curve, the velocity and the temperature evolution. We found a best fit for a kinetic energy of 0.48 foe, an initial radius of $1.8 \times 10^{13}$ cm and ejecta mass of $12.5\,{\rm M_{\odot}}$. Even including the mass for the compact remnant, this appears fully consistent with the direct measurements given above. 

\end{abstract}


\section{Introduction}

Supernovae of type IIP (SNe IIP) form a major group of core-collapse supernovae (CC SNe) 
characterized by a $3-4$ months long phase of almost constant luminosity, called the plateau. 
These events originate from the collapse of massive stars that, at the time of explosion, retain a massive hydrogen envelope. The ejecta, that is fully ionized by the initial shock breakout, cools down with the expansion of the SN and recombines powering the long plateau phase. A number of papers have shown that SNe IIP can be fruitfully used as distance indicators.
Different methods have been applied to successfully determine extragalactic distances, including the Expanding Photosphere 
Method \citep[EPM;][]{schmidt:1992, hamuy:2001, leonard:2002, dessart:2005, jones:2009}, the Spectral-fitting Expanding Atmosphere Model \citep[SEAM;][]{baron:2004} and the Standardized Candle Method 
\citep{hamuy:2002, nugent:2006, poznanski:2009, olivares:2010, dandrea:2010, maguire:2010}.
The application of all of these techniques requires well monitored light curves and high-quality spectral time series. 
Unfortunately only a limited number of type IIP SNe fulfill the required criteria.

One well-tested approach to determine the properties of the progenitors of SNe IIP is through the modeling of the SN data  with hydrodynamical codes
 \citep[e.g. ][]{grassberg:1971,falk:1977,litvinova:1983,blinnikov:2000,utrobin:2007,utrobin2:2007,utrobin:2008,pumo:2011,bersten:2012}.
The results of type IIP SN modeling have confirmed that these SNe are produced by the explosions of massive  ($\gtrsim 10\,{\rm M_{\odot}}$) red supergiant (RSG) stars.

An alternative approach for studing CC SNe is to search for their progenitors in deep pre-explosion images \citep[see][ for a review]{Sma09}. This has been done for a sample of SNe by \cite{smartt:2009}, who showed that the SNe IIP progenitor masses were confined to the range $8-17\,{\rm M}_{\odot}$, although dust effects may increase this upper limit  \citep{walmswell:2012}. 
On the other hand the theoretical expectation is that stars with masses up to 25$-$30 M$_\odot$ becames RSGs and explode as type IIP SNe. Since no star with mass above 15$-$17 M$_\odot$ has been seen to follow this path, there is an apparent discrepancy between theory and observations that has been termed the {\it Red Supergiant Problem} \citep{smartt:2009}.

However, we note that there have been claims of a discrepancy between hydrodynamical and direct progenitor mass measurements for specific SNe \citep{Sma09} that, in view of the still large uncertainties of both approaches, leave the issue still open to discussion. Therefore any opportunities for a direct comparison between these techniques are worthy of study. 

SN~2012A exploded in a very nearby host galaxy (D$<$10 Mpc) and is a first-rate target for studying the properties of the explosion and the progenitor star. This, and the fact that SN~2012A has good pre-explosion images available in public image archives, made this object as an ideal target for an extensive follow-up campaign at multiple wavelengths. In this paper we present the results of this monitoring campaign and discuss the implication for the SN progenitor star. 

The paper is organized as follows: in Section 2 we give some information about the discovery and the follow-up observations of SN~2012A, in Section 3 we present the optical and NIR photometric evolution of SN~2012A and we compare light curves, colour curves and the computed bolometric luminosity with those of other type IIP SN. In Section 4 we analyze the spectroscopic data, in Section 5 we discuss the nature of the progenitor of this SN, both by direct detection in  the pre-discovery images and by modeling of the observed data. Finally, in Sections 6 and 7 we discuss and summarize the main results of the paper. 

\section{Discovery and follow-up observations}

SN~2012A was discovered by \cite{Moore:2012}  in an image of the irregular galaxy NGC~3239 taken on 2012 Jan 7.39 UT. The discoverers reported that the transient was not visible on 2011 Dec 29 providing a useful constraint on the explosion epoch.  Spectroscopic classification obtained on Jan 10.45 UT confirmed that the transient was a type II SN close to explosion \citep{Stanishev:2012}.  Indeed a comparison  with a library of supernova spectra via the GELATO spectral classification tool \citep{avik:2008}  gives a best match with the type IIP SN~1999em  \citep{elmhamdi:2003} soon after explosion.

The SN  caught our  interest when it turned out that deep, high spatial resolution prediscovery images were available allowing for the possible identification of the SN progenitor 
\citep[] [ cf. Section~\ref{progenitor}]{Prieto:2012}. 
We immediately commenced an extensive campaign of photometric and spectroscopic monitoring that started three days after discovery and ended 
 407 days later.  

The main information on the SN and host galaxy is reported in Table~\ref{info}.

We note that the SN was also detected in $X$-rays based on {\it Swift} observations obtained in the first three weeks after explosion \citep{Pooley:2012}. 
As discussed in Section~\ref{bolometric}, it turned out that the $X$-ray detection corresponds to a negligible contribution to the bolometric flux.

\begin{table}\label{info}
\caption{Main data for SN 2012A and its host galaxy, NGC~3239 from the NASA Extragalactic Database (NED). The distance modulus \citep{mould:2000} is based on the measured redshift and using a model for the local velocity field perturbed by the influence of the Virgo  cluster, the Great Attractor (GA), and the Shapley supercluster ($H_0=73\pm5$ km s$^{-1}$ Mpc$^{-1}$). The Galactic extinction is from Schlafly \& Finkbeiner (2011). We measured a total extinction (Galactic plus host) A$_{B}$ = 0.15 mag (see Section~\ref{extinction}).
}
\begin{center}
\begin{tabular}{llll}
\hline \\
Host galaxy  & NGC 3239 \\
Galaxy type     & IB(s)m pec \\
Heliocentric velocity   & 753 $\pm$ 3  km s$^{-1}$\\
Distance modulus & 29.96 $\pm$ 0.15 mag\\
Galactic extinction A$_{B}$        & 0.117 mag\\
                                   & \\
SN type & IIP\\
RA(J2000.0) & 10$^h$25$^m$07.39$^s$\\                                   
 Dec(J2000.0) &+17$^{\circ}$09$'$14\farcs6 \\
 Offset from nucleus & 24\farcs65 E 16\farcs1 S\\
 Date of discovery & 2012 Jan 07.39 UT \\
 Estimated date of explosion & MJD=$55933.0^{+1.0}_{-3.0}$ \\
 Mag at maximum & $m_V=13.83\pm0.05$ mag\\ 
 $L_{\rm bol}$ at maximum & $1.82 \times 10^{42}$ erg s$^{-1}$ \\                                   
\hline \\
\end{tabular}
\end{center}
\label{info}
\end{table}%

\begin{figure}
\includegraphics[scale=.47,angle=0]{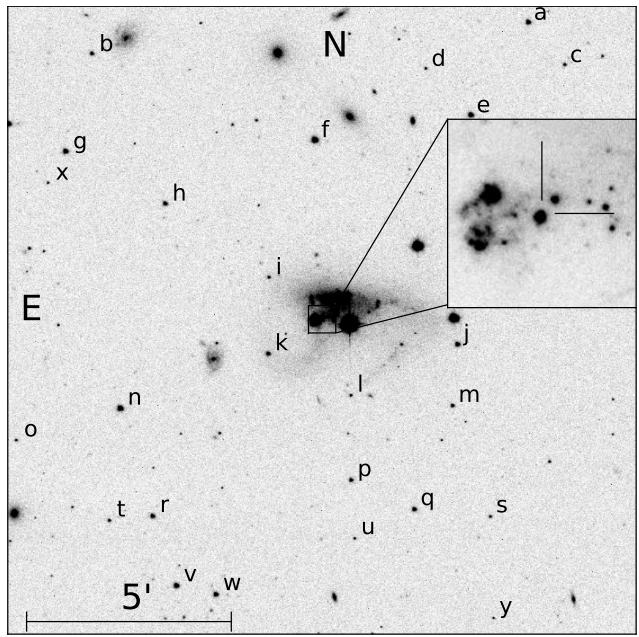}
\caption{NGC~3239 and SN~2012A. The large field $V$-band image was taken with the Schmidt 67/92 telescope, Asiago, while the zoomed view of the SN region is from the Nordic Optical Telescope, La Palma. The stars of the local sequence used for the calibration of non-photometric nights are indicated.} 
\label{map}
\end{figure}

\section{Photometry}\label{photo}

Our detailed optical and infrared photometric monitoring of SN~2012A was obtained using a large number of observing facilities, listed in Table~\ref{telescopes_phot}, where for each telescope we detail the associated instrument, the observing site, the field of view and the pixel scale. 

\begin{table*}
\caption{List of observing facilities employed for optical and infrared photometry.} \label{telescopes_phot}
\begin{tabular}{lllcc}
\hline
Telescope                        & Instrument & Site & FoV  & Scale \\
                                         &                     &        & [arcmin$^{2}$] & [arcsec pix$^{-1}$]\\
\hline
\multicolumn{5}{c}{ \bf Optical facilities }\\
Schmidt 67/92cm         & SBIG          & Asiago, Mount Ekar (Italy) &  $57\times38$ &  0.86 \\
Copernico 182cm        & AFOSC      & Asiago, Mount Ekar (Italy) &   $8 \times 8$   &  0.48 \\
Prompt   41cm              & PROMPT  & CTIO Observatory (Chile)  &    $11 \times 11$  & 0.59 \\
Calar Alto 2.2m             & CAFOS      & Calar Alto Observatory (Spain) &  $9 \times 9$ & 0.53\\
NOT     2.56m                  & ALFOSC     & Roque de los Muchachos, La Palma, Canarias (Spain) & 
$ 6.4 \times 6.4 $ & 0.19 \\
Trappist 60cm            & TRAPPISTCAM & ESO La Silla Observatory (Chile) &  $ 27 \times 27$ & 0.65 \\
ESO NTT   3.6m           & EFOSC2            & ESO La Silla Observatory (Chile) & $ 4 \times 4$  &  0.24 \\
TNG    3.6m                  &  LRS &  Roque de los Muchachos, La Palma, Canarias (Spain) &$8 \times 8 $ & 0.25 \\
\multicolumn{5}{c}{ \bf Infrared facilities }\\
REM      60cm             & REMIR & ESO La Silla Observatory (Chile) &  $10 \times 10$ & 1.22 \\
ESO NTT   3.6m           & SOFI          & ESO La Silla Observatory (Chile) & $5 \times 5 $  &  0.29 \\
NOT     2.56m              & NOTCAM     & Roque de los Muchachos, La Palma, Canarias (Spain) & $4 \times 4 $ & 0.23\\ \hline
\end{tabular}
\end{table*}

All frames were pre-processed using standard procedures in {\sc iraf} for bias subtraction, flat fielding and astrometric calibration. For infrared exposures we also applied an illumination correction and sky  background subtraction. For later epochs, multiple exposures obtained in the same night and filter were combined to improve the signal-to-noise ratio. 

To measure the SN magnitude we used a dedicated pipeline developed by one of us (EC), consisting of a collection of {\sc python} scripts calling standard {\sc iraf} tasks (through {\sc pyraf}), and  specific data analysis tools such as {\sc sextractor} for source extraction and classification, {\sc daophot} for PSF fitting and {\sc hotpants}\footnote{http://www.astro.washington.edu/users/becker/hotpants.html} for PSF matching and image subtraction.
  
Following common practice for transient photometry, as a first step we calibrated a sequence of local standards in the field. To this aim we selected among our observations those obtained on photometric nights in which standard photometric fields from the list of \cite{landolt:1992} were also observed.  These observations were used to derive the zero point and colour term for each specific instrumental set-up and to calibrate the selected stars in the SN field (see Figure~\ref{map} and Table~\ref{sequence}). The local sequence was used to calibrate non-photometric nights. For the infrared photometry we used as reference for the calibration 2MASS stars in the SN field.

\begin{table*}
\caption{Magnitudes for the local sequence stars, as indicated in Figure~\ref{map} .}\label{sequence}
\begin{tabular}{cccccccc}
\hline \\ 
id & R.A. & Dec. & $U$ & $B$ & $V$ & $R$  & $I$ \\ 
\hline \\ 
a&10:24:47.574 & 17:16:00.82 &16.371  (0.027)  &16.08  (0.003) & 15.311  (0.003) & 14.878  (0.003) & 14.480  (0.004) \\
b&10:25:29.287 & 17:15:20.22 &16.591  (0.017)  &16.64  (0.009)  &16.261  (0.006)  &16.007  (0.010)  &15.788  (0.011) \\
c&10:24:44.117 & 17:15:02.38 &18.344  (0.036)  &17.89  (0.013)  &17.003  (0.014)  &16.530  (0.021)  &16.127  (0.011) \\
d&10:24:57.404 & 17:14:58.00 &20.420  (0.117)  &19.07  (0.046)  &17.563  (0.017)  &16.601  (0.011)  &15.152  (0.016) \\
e&10:24:53.106 & 17:13:53.92 &16.224  (0.027)  &15.80  (0.005)  &15.007  (0.005)  &14.562  (0.007)  &14.186  (0.009) \\
f&10:25:08.015 & 17:13:20.67 &14.954  (0.026)  &14.74  (0.003)  &14.043  (0.004)  &13.649  (0.004)  &13.281  (0.005) \\
g&10:25:31.827 & 17:13:06.34 &15.836  (0.015)  &15.82  (0.004)  &15.129  (0.003)  &14.693  (0.004)  &14.280  (0.006) \\
h&10:25:22.336 & 17:11:54.18 &18.156  (0.027)  &17.01  (0.005)  &15.539  (0.011)  &14.601  (0.007)  &13.496  (0.010) \\
i&10:25:12.441 & 17:10:12.55 &17.353  (0.029)  &17.49  (0.021)  &17.015  (0.010)  &16.731  (0.013)  &16.429  (0.016) \\
j&10:24:54.466 & 17:08:39.59 &16.189  (0.027)  &16.04  (0.004)  &15.346  (0.006)  &14.955  (0.016)  &14.594  (0.011) \\
k&10:25:12.540 & 17:08:27.78 &18.491  (0.038)  &17.15  (0.008)  &15.821  (0.005)  &14.913  (0.011)  &14.060  (0.007) \\
l&10:25:04.672 & 17:07:30.29 &17.990  (0.033)  &18.05  (0.015)  &17.377  (0.013)  &16.985  (0.009)  &16.550  (0.034) \\
m&10:24:54.973 & 17:07:15.56 &16.708  (0.028)  &16.83  (0.008)  &16.262  (0.002)  &15.932  (0.008)  &15.594  (0.011) \\
n&10:25:26.693 & 17:07:13.71 &16.100  (0.016)  &15.47  (0.004)  &14.454  (0.005)  &13.906  (0.003)  &13.361  (0.002) \\
o&10:25:36.645 & 17:06:30.60 &20.000  (0.038)  &18.99  (0.033)  &17.481  (0.017)  &16.476  (0.023)  &15.313  (0.022) \\
p&10:25:04.685 & 17:05:34.11 &18.309  (0.036)  &17.15  (0.005)  &15.995  (0.009)  &15.198  (0.019)  &14.533  (0.007) \\
q&10:24:58.654 & 17:04:53.79 &16.692  (0.028)  &16.39  (0.007)  &15.613  (0.007)  &15.156  (0.012)  &14.732  (0.005) \\
r&10:25:23.644 & 17:04:45.78 &16.238  (0.027)  &16.18  (0.006)  &15.530  (0.009)  &15.176  (0.006)  &14.804  (0.009) \\
s&10:24:51.407 & 17:04:43.36 &17.649  (0.030)  &17.80  (0.015)  &17.206  (0.015)  &16.819  (0.018)  &16.489  (0.019) \\
t&10:25:27.785 & 17:04:39.95 &17.282  (0.029)  &17.52  (0.015)  &16.980  (0.006)  &16.679  (0.013)  &16.357  (0.019) \\
u&10:25:04.368 & 17:04:14.31 &20.092  (0.093)  &18.97  (0.032)  &17.614  (0.018)  &16.681  (0.024)  &15.880  (0.013) \\
v&10:25:21.392 & 17:03:10.35 &17.556  (0.021)  &16.34  (0.004)  &14.953  (0.009)  &14.011  (0.008)  &13.097  (0.005) \\
w&10:25:17.623 & 17:02:58.08 &16.175  (0.017)  &16.00  (0.008)  &15.293  (0.007)  &14.887  (0.004)  &14.514  (0.009) \\
x&10:25:33.513 & 17:12:23.65 &18.403  (0.029)  &18.39  (0.017)  &17.802  (0.021)  &17.436  (0.021)  &  \\
y&10:24:51.160 & 17:02:24.45 &20.088  (0.031)  &19.16  (0.019)  &18.192  (0.014)  &17.535  (0.033)  & \\
\hline \\ 
\end{tabular}
\end{table*}

The region around the SN location is crowded with many sources which require special care in the measurement of the transient instrumental magnitude, in particular when the SN becomes faint and/or the seeing is poor.   As  a rule, the SN magnitude was measured via PSF fitting. In our implementation of the PSF fitting procedure first the sky background at the SN location is estimated by a low order polynomial fit and subtracted from the image before performing a simultaneous  fit of the SN and the nearby companion stars (that is all stellar sources in a radius of $\sim 5\times$FHWM from the SN) using the PSF model derived from isolated field stars.  The fitted sources are removed from the original images, an improved estimate of the local background is derived and the PSF fitting procedure iterated. The residuals are visually inspected to validate the fit. 

An alternative approach for the measurement of transient magnitudes is template subtraction. The application of this technique requires exposures of the field obtained before the SN explosion or after the SN has faded. The template images need to be in the same filter and have good S/N and seeing. While in principle they should be obtained with the same telescope as the specific SN observation (to guarantee the same bandpass), in practice we are bound to what is actually available in the accessible image archives. In the case of SN~2012A we could retrieve deep pre-discovery exposures in $B, V, R$ bands obtained with the Vatican Advanced Telescope Technology (VATT) and $u$ and $i$ band exposures from SDSS. For the $J, H, K$ infrared bands we rely on moderate quality 2MASS exposures.

For template subtraction, first the template image is geometrically registered to the same pixel grid as the SN frame, and thereafter the PSF of the two images is matched by means of a convolution kernel determined by comparing a number of reference sources in the field. After matching the photometric scale, the template image is subtracted from the SN image and a difference image is obtained where only the transient is present. Again, the instrumental magnitude of the transient in the difference image is measured through a PSF fit, where the model PSF is derived from the image with the worst seeing. It is found that the latter procedure is more robust with respect to plain aperture photometry in particular concerning the background level determination.

When we compare the PSF fitting vs. template subtraction measured magnitudes we find that they are in excellent agreement when the transient is bright. As a rule we kept the PSF fitting magnitudes when the discrepancy is $<0.05$ mag and  instead adopted the template subtraction magnitudes when the discrepancy is larger. For the optical bands this occurs for mag $>18$ whereas for infrared the same threshold  is mag $> 14$.

Error estimates were obtained through  artificial star experiments in which a fake star, of magnitude similar to that of  the SN, is placed in the PSF fit residual image in a position close to, but not coincident with that of the real source. The simulated image is processed through the PSF fitting procedure and the dispersion of measurements out of a number of experiments, with the fake star in slightly different positions, is taken as an estimate of the instrumental magnitude error, that mainly accounts for the background fitting uncertainty. This is combined (in quadrature) with the PSF fit error returned by {\sc daophot}, and the propagated errors from the photometric calibration.

The final calibrated SN magnitudes are listed in Table~\ref{opticalphot} and \ref{sdssphot} for the optical bands and in Table~\ref{infraredphot} for the infrared.

Our light curves were complemented with $UV$-optical photometry of SN~2012A obtained using UVOT on board of the {\it Swift} satellite and recently published by \cite{Pritchard:2013}. This includes measurements in three $UV$ filters with central wavelengths in the range $1838-2600$ \AA\/, most useful for the derivation of the bolometric luminosity (cf. Section~\ref{bolometric}), as well as additional $UBV$ optical photometry.

\begin{table*}
\caption{Optical photometry in Johnson-Cousins filters, with associated errors in parentheses.}\label{opticalphot}
\begin{tabular}{cccccccc}
\hline \\ 
Date & MJD & $U$ & $B$ & $V$ & $R$  & $I$ & Instrument\\ 
\hline \\ 
20120110 & 55936.25 & 13.16 (0.07) & 14.12 (0.02) & 14.22 (0.02) & 13.89 (0.04) & 14.23 (0.06) & ALFOSC \\
20120110 & 55937.00 &  &  13.92 (0.02) &  13.94 (0.01) &  13.87 (0.02) &  13.75 (0.02) & SBIG  \\ 
20120111 & 55937.25 &  &  14.09 (0.03) &  14.03 (0.04) &  13.84 (0.03) &  13.80 (0.02) & PROMPT  \\ 
20120112 & 55938.03 &  &  13.94 (0.02) &  13.91 (0.02) &  13.79 (0.03) &  13.71 (0.03) & SBIG  \\ 
20120112 & 55938.22 &  &  14.14 (0.07) &  13.99 (0.02) &  13.90 (0.04) &  13.78 (0.04) & PROMPT  \\ 
20120112 & 55938.97 &  &  13.93 (0.04) &  13.89 (0.03) &  13.79 (0.03) &  13.71 (0.02) & SBIG  \\ 
20120113 & 55939.25 &  &  14.06 (0.03) &  14.07 (0.05) &  13.74 (0.04) &  13.72 (0.04) & PROMPT  \\ 
20120114 & 55940.03 &  &  13.86 (0.02) &  13.85 (0.02) &  13.66 (0.02) &  13.65 (0.02) & SBIG  \\ 
20120114 & 55940.26 &  &  13.94 (0.03) &  14.01 (0.01) &  13.82 (0.02) &  13.73 (0.02) & PROMPT  \\ 
20120114 & 55940.92 &  &  13.92 (0.02) &  13.84 (0.02) &  13.68 (0.03) &  13.69 (0.02) & SBIG  \\ 
20120115 & 55941.97 &  &  13.96 (0.03) &  13.83 (0.02) &  13.71 (0.03) &  13.68 (0.02) & SBIG  \\ 
20120118 & 55944.05 &  &  13.98 (0.04) &  13.85 (0.04) &  13.67 (0.03) &  13.64 (0.03) & SBIG  \\ 
20120121 & 55947.22 &  &  14.20 (0.02) &  13.96 (0.01) &  13.71 (0.05) &  13.61 (0.02) & PROMPT  \\ 
20120121 & 55947.93 & 13.85 (0.05) &  14.26 (0.02) &  13.97 (0.01) &  13.75 (0.03) &  13.70 (0.07) & AFOSC  \\ 
20120123 & 55949.96 & 14.26 (0.03) &  14.39 (0.02) &  14.01 (0.05) &  13.69 (0.03) &  13.47 (0.04) & AFOSC  \\ 
20120124 & 55950.93 & 14.26 (0.02) &  14.33 (0.01) &  13.91 (0.04) &  13.60 (0.02) &  13.52 (0.05) & AFOSC  \\ 
20120125 & 55951.21 &  &  14.48 (0.26) &   &   &   & PROMPT  \\ 
20120125 & 55951.93 &  &  14.37 (0.03) &  13.87 (0.02) &  13.63 (0.04) &  13.57 (0.03) & SBIG  \\ 
20120126 & 55952.23 &  &   &  14.05 (0.08) &  13.78 (0.14) &  13.61 (0.02) & PROMPT  \\ 
20120127 & 55953.05 & 14.56 (0.04) &  14.61 (0.05) &  14.06 (0.03) &  13.56 (0.03) &  13.52 (0.04) & AFOSC  \\ 
20120127 & 55953.88 & 14.63 (0.07) &  14.65 (0.09) &  14.11 (0.03) &  13.65 (0.05) &  13.55 (0.12) & AFOSC  \\ 
20120129 & 55955.94 & 15.00 (0.03) &  14.75 (0.04) &  14.11 (0.02) &  13.71 (0.02) &  13.63 (0.04) & AFOSC  \\ 
20120130 & 55956.93 & 15.38 (0.03) &  14.81 (0.02) &  14.13 (0.03) &  13.79 (0.04) &  13.68 (0.06) & AFOSC  \\ 
20120203 & 55960.93 &  &   &  14.09 (0.04) &  13.71 (0.04) &  13.67 (0.04) & SBIG  \\ 
20120208 & 55965.15 &  &   &  14.26 (0.10) &  13.86 (0.06) &   & SBIG  \\ 
20120208 & 55965.93 &  &  15.02 (0.04) &  14.15 (0.03) &  13.76 (0.07) &  13.61 (0.05) & SBIG  \\ 
20120209 & 55966.19 &  &  15.11 (0.04) &  14.27 (0.05) &  13.82 (0.05) &  13.60 (0.04) & PROMPT  \\ 
20120211 & 55968.08 & 15.63 (0.03) &  15.20 (0.05) &  14.34 (0.01) &  13.86 (0.02) &  13.62 (0.02) & CAFOS  \\ 
20120214 & 55971.18 &  &  15.19 (0.03) &  14.30 (0.03) &  13.93 (0.04) &  13.67 (0.03) & PROMPT  \\ 
20120217 & 55974.17 &  &  15.34 (0.04) &  14.38 (0.02) &  13.89 (0.03) &  13.60 (0.03) & PROMPT  \\ 
20120217 & 55974.97 & 16.44 (0.04) &  15.35 (0.02) &  14.29 (0.02) &  13.87 (0.05) &   & AFOSC  \\ 
20120218 & 55975.97 & 16.28 (0.06) &  15.50 (0.02) &  14.43 (0.03) &  13.82 (0.04) &  13.66 (0.06) & AFOSC  \\ 
20120221 & 55978.31 &  &   &  14.38 (0.02) &  13.93 (0.04) &  13.66 (0.03) & PROMPT  \\ 
20120223 & 55980.04 & 16.45 (0.07) &  15.53 (0.03) &  14.42 (0.03) &  14.00 (0.04) &  13.68 (0.05) & AFOSC  \\ 
20120224 & 55981.20 &  &  15.46 (0.03) &  14.41 (0.04) &  13.96 (0.03) &  13.68 (0.03) & PROMPT  \\ 
20120226 & 55983.01 & 16.66 (0.04) &  15.50 (0.02) &  14.42 (0.02) &  13.81 (0.02) &  13.80 (0.04) & AFOSC  \\ 
20120227 & 55984.14 &  &  15.53 (0.05) &  14.41 (0.02) &  13.99 (0.03) &  13.68 (0.04) & PROMPT  \\ 
20120228 & 55985.96 & 16.93 (0.06) &  15.62 (0.03) &  14.50 (0.04) &  13.97 (0.02) &  13.75 (0.05) & AFOSC  \\ 
20120229 & 55986.14 &  &  15.47 (0.14) &   &   &   & PROMPT  \\ 
20120301 & 55987.14 &  &   &  14.48 (0.02) &  14.03 (0.04) &  13.74 (0.03) & PROMPT  \\ 
20120304 & 55990.12 &  &  15.65 (0.03) &  14.49 (0.03) &  14.02 (0.03) &  13.66 (0.02) & PROMPT  \\ 
20120306 & 55992.13 &  &  15.57 (0.17) &   &   &   & PROMPT  \\ 
20120306 & 55992.89 &  &   &  14.45 (0.03) &  13.93 (0.03) &  13.76 (0.04) & SBIG  \\ 
20120312 & 55998.07 &  &  15.91 (0.02) &  14.72 (0.03) &  14.09 (0.02) &  13.84 (0.04) & CAFOS  \\ 
20120312 & 55998.11 &  &  15.73 (0.05) &  14.57 (0.01) &  14.11 (0.02) &  13.78 (0.02) & PROMPT  \\ 
20120313 & 55999.16 & 17.19 (0.06) &  15.72 (0.02) &  14.69 (0.04) &  14.12 (0.01) &  14.04 (0.05) & EFOSC  \\ 
20120313 & 55999.80 &  &  15.70 (0.03) &  14.60 (0.02) &  14.10 (0.03) &  13.87 (0.02) & SBIG  \\ 
20120314 & 56000.10 &  &  15.78 (0.03) &  14.59 (0.03) &  14.10 (0.04) &  13.83 (0.02) & PROMPT  \\ 
20120314 & 56000.85 &  &  15.65 (0.02) &  14.51 (0.02) &  14.07 (0.02) &  13.78 (0.02) & SBIG  \\ 
20120315 & 56001.79 &  &  15.87 (0.02) &  14.58 (0.01) &  14.20 (0.03) &  13.91 (0.02) & SBIG  \\ 
20120316 & 56002.05 &  &  15.64 (0.21) &   &   &   & PROMPT  \\ 
20120316 & 56002.82 &  &  15.65 (0.02) &  14.52 (0.02) &  14.15 (0.02) &  13.79 (0.03) & SBIG  \\ 
20120317 & 56003.04 &  &   &  14.71 (0.04) &  14.17 (0.03) &  13.95 (0.03) & PROMPT  \\ 
20120317 & 56003.96 & 17.23 (0.08) &  15.81 (0.03) &  14.55 (0.02) &  14.07 (0.04) &   & AFOSC  \\ 
20120319 & 56005.12 &  &  15.83 (0.04) &  14.67 (0.03) &  14.19 (0.03) &  13.92 (0.03) & PROMPT  \\ 
20120320 & 56006.88 &  &  15.73 (0.05) &  14.52 (0.02) &  14.16 (0.04) &  13.76 (0.04) & SBIG  \\ 
\hline \\ 
\end{tabular}
\end{table*}

\begin{table*}
\contcaption{Optical photometry in Johnson-Cousins filters, with associated errors in parentheses.}
\begin{tabular}{cccccccc}
\hline \\ 
Date & MJD & $U$ & $B$ & $V$ & $R$  & $I$ & Instrument \\ 
\hline \\ 
20120322 & 56008.11 &  &  15.92 (0.03) &  14.70 (0.05) &  14.19 (0.02) &  13.96 (0.02) & PROMPT  \\ 
20120322 & 56008.83 &  &  15.87 (0.03) &  14.66 (0.02) &  14.24 (0.03) &  14.04 (0.03) & SBIG  \\ 
20120324 & 56010.07 &  &  15.93 (0.04) &  14.71 (0.03) &  14.25 (0.03) &  13.94 (0.01) & PROMPT  \\ 
20120326 & 56012.08 &  &  15.83 (0.14) &   &   &   & PROMPT  \\ 
20120326 & 56012.84 &  &  15.89 (0.05) &  14.69 (0.01) &  14.28 (0.02) &  13.99 (0.03) & SBIG  \\ 
20120327 & 56013.01 &  &   &  14.90 (0.03) &  14.35 (0.03) &  14.07 (0.02) & PROMPT  \\ 
20120327 & 56013.86 &  &  15.93 (0.03) &  14.71 (0.02) &  14.43 (0.07) &   & SBIG  \\ 
20120328 & 56014.82 &  &  15.95 (0.05) &  14.77 (0.04) &  14.26 (0.05) &  14.08 (0.04) & SBIG  \\ 
20120329 & 56015.81 &  &  15.90 (0.03) &  14.82 (0.02) &  14.37 (0.02) &  14.10 (0.03) & SBIG  \\ 
20120330 & 56016.05 &  &  16.02 (0.03) &  14.87 (0.03) &  14.36 (0.03) &  14.06 (0.02) & PROMPT  \\ 
20120331 & 56017.94 & 17.83 (0.17) &   &   &   &   & AFOSC  \\ 
20120331 & 56017.94 &  &  16.16 (0.04) &  15.03 (0.03) &  14.53 (0.01) &  14.16 (0.05) & AFOSC  \\ 
20120403 & 56020.04 &  &  16.23 (0.05) &  14.96 (0.02) &  14.48 (0.03) &  14.14 (0.03) & PROMPT  \\ 
20120409 & 56026.11 &  &  16.37 (0.07) &  15.17 (0.03) &  14.66 (0.03) &  14.35 (0.01) & PROMPT  \\ 
20120412 & 56029.08 &  &   &   &  14.80 (0.23) &  14.49 (0.09) & PROMPT  \\ 
20120412 & 56030.00 &  &  16.53 (0.02) &  15.31 (0.01) &  14.81 (0.03) &  14.57 (0.02) & ALFOSC  \\ 
20120414 & 56032.00 &  &  16.69 (0.02) &  15.39 (0.02) &  14.99 (0.02) &  14.56 (0.05) & ALFOSC  \\ 
20120417 & 56034.11 &  &   &  15.64 (0.03) &  14.97 (0.03) &  14.72 (0.02) & PROMPT  \\ 
20120419 & 56036.00 &  &  17.13 (0.05) &  15.80 (0.05) &  15.13 (0.03) &  14.90 (0.02) & PROMPT  \\ 
20120421 & 56039.00 &  &  17.62 (0.10) &  16.26 (0.05) &  15.52 (0.04) &  15.18 (0.03) & PROMPT  \\ 
20120422 & 56039.90 &  &   &  16.34 (0.03) &  15.49 (0.04) &  15.41 (0.11) & SBIG  \\ 
20120423 & 56040.04 & $>19.40$  &  17.87 (0.04) &  16.40 (0.03) &  15.67 (0.04) &  15.26 (0.03) & CAFOS  \\ 
20120424 & 56041.96 &  $>18.90$ &  18.32 (0.14) &  16.79 (0.05) &  15.95 (0.04) &   & AFOSC  \\ 
20120425 & 56042.86 &  &  18.28 (0.05) &  16.77 (0.03) &  16.09 (0.03) &  15.70 (0.03) & SBIG  \\ 
20120426 & 56043.86 &  &  18.64 (0.07) &  17.13 (0.04) &  16.22 (0.05) &  15.89 (0.04) & SBIG  \\ 
20120427 & 56044.82 &  &  18.58 (0.09) &  17.23 (0.03) &  16.34 (0.05) &  15.90 (0.05) & SBIG  \\ 
20120501 & 56048.85 &  &  18.90 (0.07) &  17.63 (0.04) &  16.61 (0.04) &  16.19 (0.04) & CAFOS  \\ 
20120515 & 56062.86 &  &  19.16 (0.15) &  17.72 (0.02) &  16.76 (0.03) &  16.32 (0.06) & CAFOS  \\ 
20120529 & 56077.00 &  &  19.40 (0.11) &  18.02 (0.10) &  17.01 (0.10) &   & TRAPPISTCAM  \\ 
20120619 & 56097.99 &  &  19.48 (0.10) &  18.03 (0.10) &  17.15 (0.11) &   & TRAPPISTCAM \\
20121020 & 56220.18 &  &  20.12 (0.15) &  18.99 (0.13) &   &   & AFOSC  \\ 
20121021 & 56221.25 &   &  20.31 (0.15)  &   19.11 (0.14) & 18.03  (0.03) &  17.83 (0.03)  & LRS \\
20121109 & 56240.17 &  &  20.54 (0.27) &  19.37 (0.14) &  18.19 (0.09) &  17.95 (0.23) & AFOSC  \\ 
20121122 & 56253.34 &  &                         &  19.52 (0.11) &  18.41 (0.12)  &  & TRAPPISTCAM \\
20121207 & 56268.13 &  &  20.60 (0.23) &  19.45 (0.15) &  18.65 (0.10) &  18.25 (0.14) & AFOSC  \\ 
20121210 & 56271.15 &  &   &  19.55 (0.14) &  18.63 (0.11) &  18.45 (0.12) & AFOSC  \\ 
20121212 & 56273.28  &  &                        &                         &  18.67 (0.11)  &  & TRAPPISTCAM \\
20130203 & 56326.25 &  &  20.86 (0.15)  &   19.96 (0.14) & 19.09  (0.14) &  19.07 (0.11)  & LRS \\
20130221 & 56344.24  & &                          &   20.07 (0.09) & 19.39 (0.10) &                           &   TRAPPISTCAM \\\hline \\ 
\end{tabular}
\end{table*}

\begin{table*}
\caption{Infrared photometry calibrated to the 2MASS system, with errors in parentheses.}\label{infraredphot}
\begin{tabular}{cccccc}
\hline \\ 
Date & MJD & $J$ & $H$ & $K$  & Instrument\\ 
\hline \\ 
20120111 & 55937.23 & 14.15 (0.16) &  13.72 (0.17) &  13.55 (0.18) & REMIR  \\ 
20120112 & 55938.29 & 14.14 (0.15) &  13.71 (0.30) &  13.55 (0.26) & REMIR  \\ 
20120113 & 55939.35 & 13.61 (0.16) &  13.89 (0.17) &  13.44 (0.33) & REMIR  \\ 
20120124 & 55950.26 & 13.43 (0.32) &  13.38 (0.26) &  13.31 (0.39) & REMIR  \\ 
20120211 & 55968.12 & 13.38 (0.11) &  13.04 (0.14) &  13.35 (0.13) & REMIR  \\ 
20120216 & 55973.29 & 13.37 (0.46) &  13.09 (0.40) &  13.22 (0.32) & REMIR  \\ 
20120225 & 55982.08 & 13.28 (0.29) &  12.97 (0.23) &  12.98 (0.28) & REMIR  \\ 
20120305 & 55991.09 & 13.67 (0.17) &  13.66 (0.19) &  13.29 (0.39) & REMIR  \\ 
20120314 & 56000.08 & 13.56 (0.12) &  13.34 (0.17) &  13.30 (0.28) & SOFI  \\ 
20120315 & 56001.18 & 13.68 (0.25) &  13.31 (0.16) &  13.15 (0.14) & SOFI  \\ 
20120325 & 56011.01 & 13.83 (0.18) &  13.32 (0.06) &  13.47 (0.14) & REMIR  \\ 
20120405 & 56022.11 & 13.78 (0.04) &  13.70 (0.06) &  13.65 (0.20) & REMIR  \\ 
20120412 & 56029.11 & 14.18 (0.19) &  13.92 (0.08) &  14.11 (0.16) & REMIR  \\ 
20120416 & 56033.10 & 14.34 (0.23) &  14.01 (0.07) &  14.14 (0.08) & REMIR  \\ 
20120421 & 56038.10 & 14.66 (0.07) &  14.47 (0.10) &  14.30 (0.14) & REMIR  \\ 
%20120430 & 56047.08 & 15.85 (0.18) &  15.59 (0.22) &  15.26 (0.21) & REMIR  \\ 
20120507 & 56054.00 & 15.35 (0.29) &  14.93 (0.12) &  14.91 (0.16) & NOTCAM  \\ 
\hline \\ 
\end{tabular}
\end{table*}

\subsection{Photometric evolution}\label{photo_evol}

\begin{figure}
\includegraphics[scale=.43,angle=0]{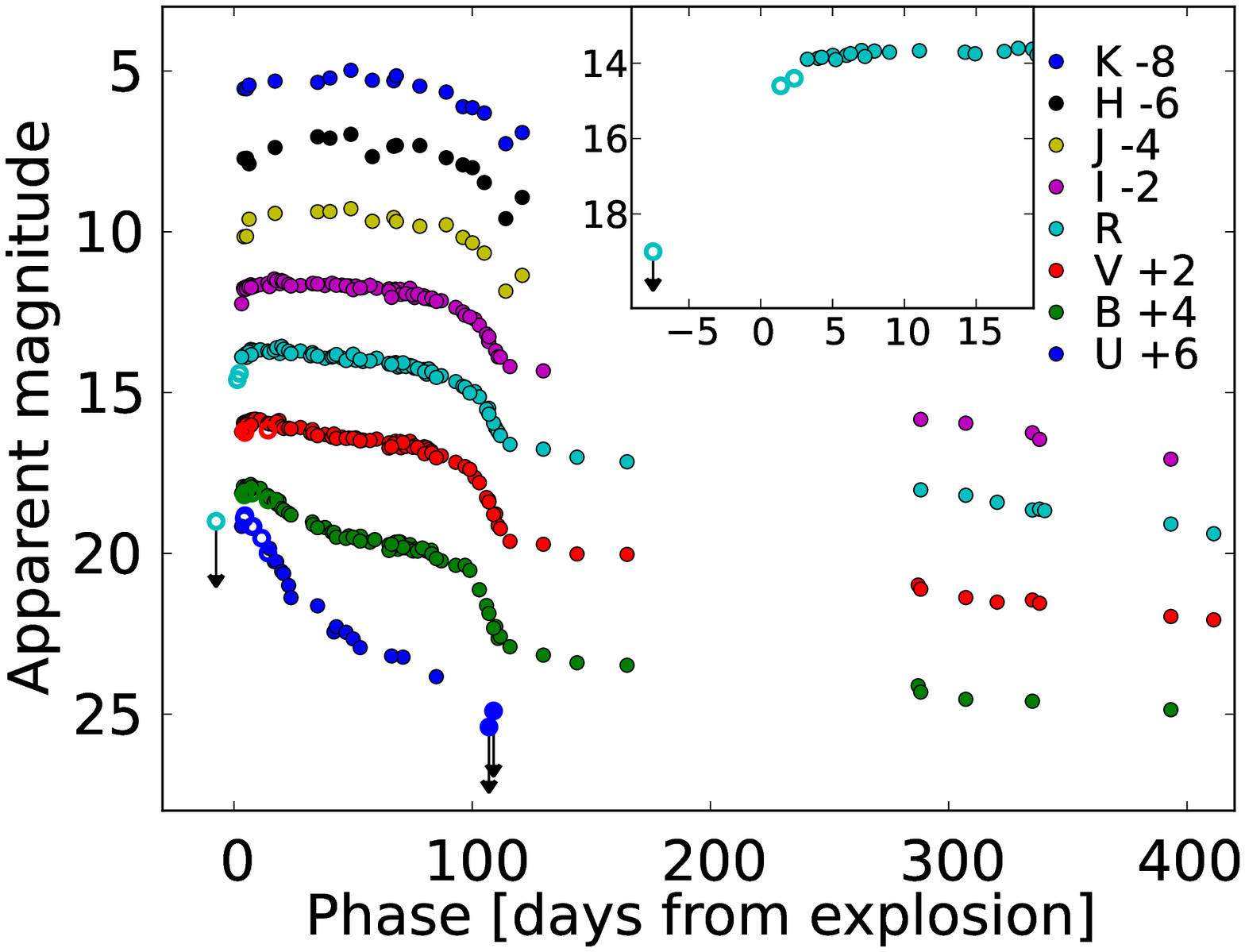}
\caption{Light curves of SN 2012A in $UBVRIJHK$ bands. The insert is a zoom on the early $R$-band light curve.} 
\label{lightcurve}
\end{figure}

The multicolour light curves of SN 2012A  are shown in Figure~\ref{lightcurve}. We included the pre-discovery limit and the discovery and confirmation magnitudes  from \cite{Moore:2012} that, although unfiltered, are crucial to constrain the epoch of explosion. Indeed, when compared with our $R$-band photometry, these observations appear to describe a very steep rise to maximum and indicate MJD=$55933.0^{+1.0}_{-3.0}$  as the best estimate for the epoch of explosion.

In  typical SNe~IIP, after the fast rise, the light curve settles on a plateau phase during which the luminosity remains roughly constant, sustained by the recombination of the hydrogen envelope which was fully ionized after shock breakout. Actually, in the case of SN~2012A the luminosity in the plateau phase was never really constant:  between 20 and 80 days, the magnitude decline rate is indeed small in the $K$-band [$0.3~{\rm mag}\,(100{\rm d})^{-1}$], but already significant in the $R$-band [$1.0~{\rm mag}\,(100{\rm d})^{-1}$] and is largest in the $U$-band
[$4.0~{\rm mag}\,(100{\rm d})^{-1}$].  For comparison, in the same period SN~1999em \citep[cf. ][]{elmhamdi:2003} brightened in the $K$-band at a rate of  $1.0~{\rm mag}\,(100{\rm d})^{-1}$, declined by only $0.1~{\rm mag}\,(100{\rm d})^{-1}$ in the $R$-band and had a similar decline in the $U$-band of $5.0~{\rm mag}\,(100{\rm d})^{-1}$.

We note that the behavior in the different bands of SN~2012A does not appear to be driven by a decrease of the photospheric temperature in the envelope that, as expected in the recombination phase, remains more or less constant (cf. Figure~\ref{tbb}). Most likely the evolution in the plateau phase is related to changes in line opacities in the outer envelope regions.

At the end of hydrogen recombination the light curves of type IIP SNe show a sudden drop in luminosity.  In the case of SN~2012A this occurs about 90 days after explosion, $15-20$ days earlier than for SN~1999em. The period of rapid luminosity decline terminates at about $110$ days after a drop in the $R$-band of $\sim$ 2 mag. After this,  the light curves settle onto a much shallower linear decline powered  by the radioactive decay of  $^{56}$Co into $^{56}$Fe. Indeed, the decline rate in $V$ and $R$ bands measured in the phase range $150-400\,{\rm d}$ is $0.9~{\rm mag}\,(100{\rm d})^{-1}$, very close to the expected energy input from $^{56}$Co decay, $0.98~{\rm mag}\,(100{\rm d})^{-1}$.

\begin{figure}
\includegraphics[scale=.5,angle=0]{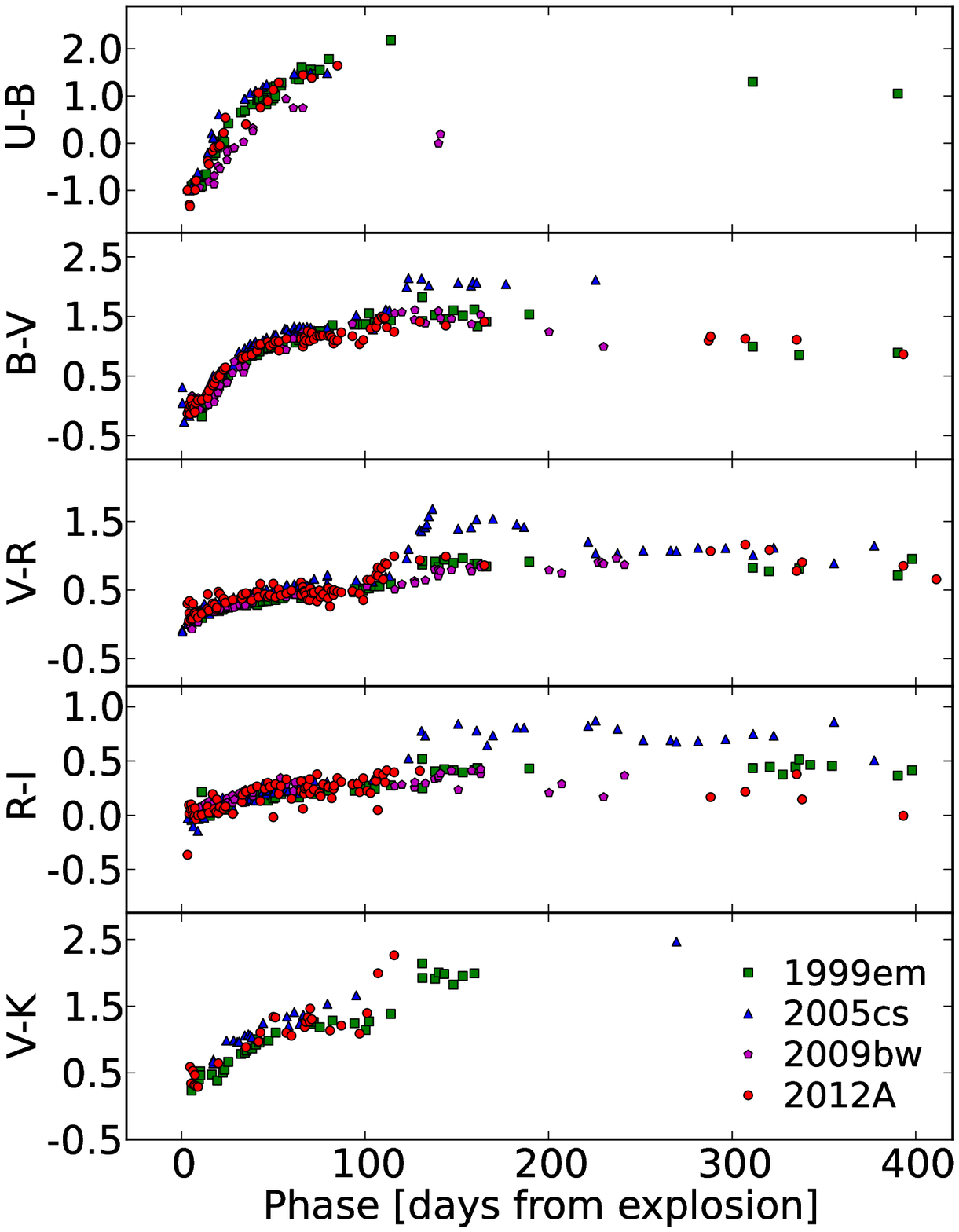}
\caption{From the top to the bottom: $U-B$, $B - V$, $V - R$, $R - I$, $V - K$ colours of SN 2012A from early times to the nebular phase, compared to SNe 2005cs, 1999em and 2009bw (see text for discussion).} 
\label{col}
\end{figure}

The $U-B$, $B-V$, $V-R$, $R-I$ and $V-K$ colour curves of SN~2012A are shown in Figure~\ref{col} and compared to those of SNe 1999em  \citep{elmhamdi:2003}, 2009bw \citep{inserra:2012}  and 2005cs \citep{pastorello:2006,pastorello:2009}. 
The choice of these SNe as references is justified by the fact that, as will be shown in the following, they appear to bracket SN~2012A in a number of properties, including luminosity, expansion velocity, temperature and $^{56}$Ni-mass.

The colours of SN~2012A have been corrected for the sum of the  adopted Galactic and host galaxy extinction, that is $A_B=0.15$ mag (cf. Section~\ref{extinction}), whereas for the comparison SNe we adopted the extinction corrections given in the quoted papers. 
In all colours the evolution of the four SNe is remarkably similar up to 100 days from explosion.
The rapid colour evolution seen in the first month (especially in the $B-V$ colour) is a result of the expansion and cooling of the SN photosphere. After $100-120$ days, the colours of SN~2012A show very little evolution, similar to the behaviour of SNe~1999em and 2009bw,  whereas SN 2005cs shows a sudden jump to somewhat redder colours that is a characteristic feature of many faint type II SNe \citep{pastorello:2009}. 

\subsection{The {\em ugriz} photometry}

An increasing number of observing facilities are equipped with {\em ugriz} filters which mimic the photometric system of the Sloan Digital Sky Survey\footnote{http://www.sdss.org}.  Therefore one is often faced with the problem of comparing SDSS-like photometry for a SN with the template light curves available in the conventional {\em UBVRI} Johnson-Cousins system. Transformation equations 
have been provided by a number of authors, derived from comparisons of the magnitudes of standard stars in the two photometric systems \citep[][and reference therein]{chonis:2008}. However SN spectra are very different from those of typical stars, with strong absorptions and emissions, and so the validity of these transformation equations is not guaranteed.

For SN~2012A, in many respects a typical type IIP SN, we have the opportunity of direct testing the validity of these transformation equations, thanks to simultaneous observations in both systems obtained with the PROMPT telescopes at CTIO. 

The {\em griz} photometry of SN~2012A, reduced with the same recipes as for the Johnson-Cousins photometry, is reported in Table~\ref{sdssphot}. The photometric calibration was obtained by comparison with the SDSS magnitudes for the stars in the field of the SN and are therefore close to the AB system ($SDSS = AB - 0.02$ mag), whereas the Johnson-Cousins photometry reported in Table~\ref{opticalphot} are in the VEGA system.

\begin{figure}
\includegraphics[scale=.5,angle=0]{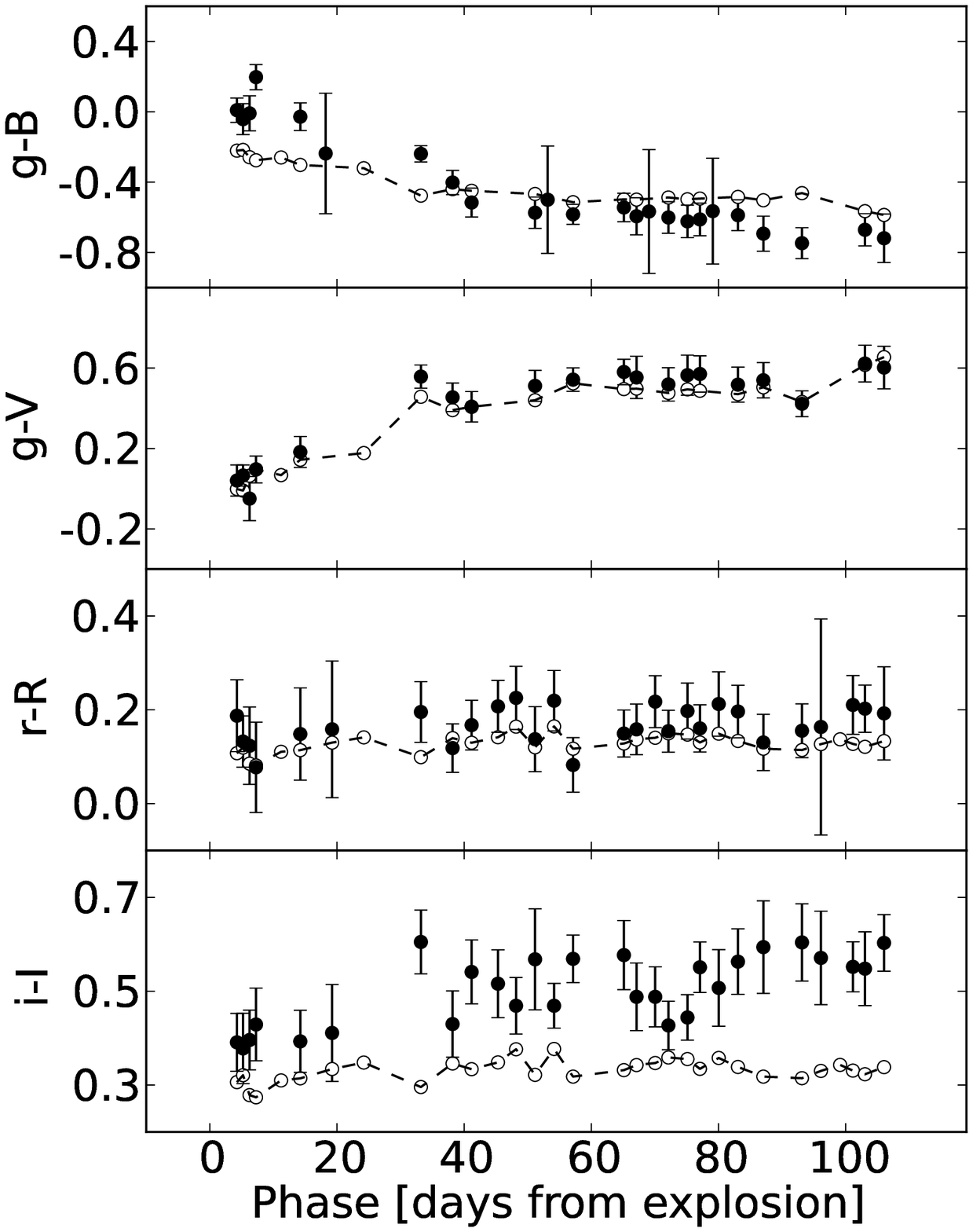}
\caption{Comparison between $griz$ and Johnson-Cousins photometry for SN~2012A. The differences between the observed magnitudes (filled circles) is compared with the  expected differences (empty circles) as derived from the transformation equations of Chonis \& Gaskell (2008), see text for discussion. } 
\label{sloan}
\end{figure}

In Figure~\ref{sloan} we show the difference in SN magnitude in the SDSS and Johnson-Cousins systems for selected filter combinations as a function of the light curve phase. The difference between the observed magnitudes (filled circles) is compared with the  expected difference (empty circles) as derived from the transformation equations of \cite{chonis:2008} using the observed $g-r$ and $r-i$ colours. 

Besides the $AB-VEGA$ systematic offset, Figure~\ref{sloan} shows in all bands apart from $r-R$ a clear trend with phase. This is largely expected because of the rapid SN colour evolution and dependence of the transformation equation on the colour. Indeed it appears that the  transformation equations are a fairly good approximation for the $g-V$ and $r-R$ differences, but less valid for the $g-B$ and $i-I$ differences. While the differences in $g-B$ are readily attributed to the very different passbands of the two filters resulting in a different sampling of the evolving spectral energy distribution of the SN, the large systematic offset in $i-I$ required further analysis. It transpired that the $i$ filter of PROMPT has a somewhat redder cutoff than the SDSS $i$ filter and hence it includes the strong Ca~II feature at $\sim 8600$ \AA\/ that is outside of the SDSS $i$ band. This effect accounts for the offset of $\sim 0.15$ mag of the observed differences and those expected from the transformation equations.

We conclude that provided filters are carefully matched, transformation equations
can be safely used for SNe IIP, apart from for $g-B$.  

\begin{table*}
\caption{Photometry in {\em griz} bands. The magnitude system is as for SDSS DR7, that is $SDSS = AB - 0.02$ mag. Errors are given in parentheses.}\label{sdssphot}
\begin{tabular}{ccccccc}
\hline \\ 
Date & MJD & $g$ & $r$ & $i$  & $z$ & Instrument\\ 
\hline \\ 
20120113 & 55939.25 & 14.04 ( 0.10) &  13.88 ( 0.07) &  14.12 ( 0.05) &  14.30 ( 0.05) & PROMPT  \\ 
20120114 & 55940.27 & 14.13 ( 0.07) &  13.90 ( 0.10) &  14.17 ( 0.07) &  14.23 ( 0.05) & PROMPT  \\ 
20120118 & 55944.21 & 14.07 ( 0.07) &  13.90 ( 0.04) &  14.06 ( 0.04) &  14.25 ( 0.10) & PROMPT  \\ 
20120121 & 55947.23 & 14.16 ( 0.08) &  13.86 ( 0.08) &  14.01 ( 0.06) &  14.15 ( 0.08) & PROMPT  \\ 
20120125 & 55951.22 & 14.23 ( 0.22) &   &   &   & PROMPT  \\ 
20120126 & 55952.24 &  &  13.95 ( 0.05) &  14.03 ( 0.10) &  14.16 ( 0.10) & PROMPT  \\ 
20120131 & 55957.21 & 14.45 ( 0.07) &  14.09 ( 0.06) &  14.13 ( 0.08) &  14.01 ( 0.07) & PROMPT  \\ 
20120209 & 55966.20 & 14.86 ( 0.03) &  14.02 ( 0.04) &  14.22 ( 0.06) &  14.01 ( 0.06) & PROMPT  \\ 
20120214 & 55971.18 & 14.78 ( 0.06) &  14.06 ( 0.03) &  14.11 ( 0.07) &  14.14 ( 0.09) & PROMPT  \\ 
20120217 & 55974.18 & 14.82 ( 0.07) &  14.07 ( 0.04) &  14.15 ( 0.06) &  14.03 ( 0.04) & PROMPT  \\ 
20120221 & 55978.32 &  &  14.14 ( 0.04) &  14.18 ( 0.07) &  14.14 ( 0.07) & PROMPT  \\ 
20120224 & 55981.17 &  &  14.20 ( 0.06) &  14.15 ( 0.06) &  14.13 ( 0.07) & PROMPT  \\ 
20120227 & 55984.15 & 14.95 ( 0.07) &  14.14 ( 0.06) &  14.26 ( 0.10) &  13.99 ( 0.08) & PROMPT  \\ 
20120229 & 55986.14 & 14.96 ( 0.27) &   &   &   & PROMPT  \\ 
20120301 & 55987.14 &  &  14.26 ( 0.05) &  14.21 ( 0.04) &  14.19 ( 0.06) & PROMPT  \\ 
20120304 & 55990.13 & 15.06 ( 0.05) &  14.11 ( 0.05) &  14.24 ( 0.05) &  14.05 ( 0.06) & PROMPT  \\ 
20120312 & 55998.12 & 15.17 ( 0.06) &  14.27 ( 0.04) &  14.36 ( 0.07) &  14.12 ( 0.06) & PROMPT  \\ 
20120314 & 56000.11 & 15.17 ( 0.10) &  14.27 ( 0.04) &  14.32 ( 0.07) &  14.13 ( 0.06) & PROMPT  \\ 
20120316 & 56002.07 & 15.06 ( 0.28) &   &   &   & PROMPT  \\ 
20120317 & 56003.05 &  &  14.40 ( 0.05) &  14.44 ( 0.06) &  14.24 ( 0.11) & PROMPT  \\ 
20120319 & 56005.12 & 15.22 ( 0.08) &  14.35 ( 0.03) &  14.36 ( 0.04) &  14.25 ( 0.05) & PROMPT  \\ 
20120322 & 56008.12 & 15.29 ( 0.09) &  14.39 ( 0.06) &  14.41 ( 0.04) &  14.30 ( 0.05) & PROMPT  \\ 
20120324 & 56010.08 & 15.30 ( 0.09) &  14.42 ( 0.04) &  14.50 ( 0.05) &  14.34 ( 0.10) & PROMPT  \\ 
20120326 & 56012.10 & 15.26 ( 0.27) &   &   &   & PROMPT  \\ 
20120327 & 56013.04 &  &  14.57 ( 0.06) &  14.58 ( 0.08) &  14.48 ( 0.07) & PROMPT  \\ 
20120330 & 56016.07 & 15.42 ( 0.08) &  14.56 ( 0.05) &  14.63 ( 0.07) &  14.51 ( 0.06) & PROMPT  \\ 
20120403 & 56020.06 & 15.53 ( 0.09) &  14.61 ( 0.05) &  14.74 ( 0.10) &  14.53 ( 0.09) & PROMPT  \\ 
20120409 & 56026.13 & 15.61 ( 0.06) &  14.82 ( 0.05) &  14.96 ( 0.08) &  14.64 ( 0.13) & PROMPT  \\ 
20120412 & 56029.11 &  &  14.97 ( 0.04) &  15.07 ( 0.04) &  14.79 ( 0.11) & PROMPT  \\ 
20120415 & 56032.13 &  &  15.10 ( 0.07) &  15.16 ( 0.07) &  14.94 ( 0.09) & PROMPT  \\ 
20120417 & 56034.15 &  &  15.19 ( 0.06) &  15.28 ( 0.05) &  15.05 ( 0.10) & PROMPT  \\ 
20120419 & 56036.04 & 16.45 ( 0.07) &  15.34 ( 0.04) &  15.46 ( 0.08) &  15.21 ( 0.08) & PROMPT  \\ 
20120422 & 56039.04 & 16.89 ( 0.09) &  15.72 ( 0.09) &  15.79 ( 0.05) &  15.42 ( 0.22) & PROMPT  \\ 
\hline \\ 
\end{tabular}
\end{table*}

\subsection{Distance and absolute magnitudes}\label{distance}

The {\em Nearby Galaxies Catalog} by \cite{Tully:1988} reports for NGC~3239 a distance of 8.1 Mpc, based on the measured redshift and using a model for the local velocity field.
Indeed, as is well known, the redshift of nearby galaxies is perturbed by the influence of the Virgo  cluster, the Great Attractor (GA), and the Shapley supercluster. It turns out that updated models of the local velocity field yield a distance that is about 20\% larger than the above value, that is $9.8 \pm 0.7$ Mpc as also reported by the NASA/IPAC Extragalactic database (NED)\footnote{http://ned.ipac.caltech.edu}. This is fully consistent with the value of $10.0$ Mpc reported in the {\em Extragalactic Distance Database} \citep{Tully:2009} and obtained with the same method.  In the following we adopt a distance of 9.8 Mpc, i.e. a distance modulus of 29.96 mag with a formal error of 0.15 mag as reported by NED ($H_0=73$ km s$^{-1}$ Mpc$^{-1}$). 
We also derive the total extinction in the direction of NGC~3239 (Galactic plus host galaxy) as $A_B=0.15$ mag  (see Section~\ref{extinction}).

The SN apparent magnitudes at maximum were estimated through a low order polynomial fit of the early light curves ($<$ 30 d), from which we obtained $m_B=13.98\pm0.03$ mag, $m_V=13.81\pm0.03$ mag and $m_R=13.66\pm0.03$ mag, where the error budget is dominated by the fact that, with the exception of the $R$-band, the rise to the maximum light is not really well constrained. Keeping in mind this uncertainty in the detection of maximum epoch, we derived SN absolute magnitudes at maximum of $M_B=-16.23\pm0.16$ mag, $M_V=-16.28\pm0.16$ mag, $M_R=-16.41\pm0.16$ mag, close to the average for SNe~IIP \citep{Li:2011}.  In particular at maximum SN~2012A was very similar to
SN 1999em which had $M_B=-16.1\pm0.4$ mag (with a relatively large uncertainty in the host galaxy distance) and about 1 mag brighter than the faint type IIP SN 2005cs with  $M_B=-15.1\pm0.3$ mag. 

We can use the \cite{nugent:2006} correlation between the absolute brightness of SNe~IIP and the expansion velocities derived from the minimum of the 
Fe~II 5169 \AA\/ P-Cygni feature observed during the plateau phase. Incorporating our $V-I$ colour and the velocity measured from the Fe~II 5169 \AA\/ line at +50~d in equation~(1) of \cite{nugent:2006}, we find a distance modulus of $29.72\pm0.17$ mag which, within the uncertainties, is consistent with the assumed value.

\subsection{Bolometric light curve}\label{bolometric}

By integrating the multicolour photometry of SN~2012A from the $UV$ to the near $IR$, we can derive the bolometric luminosity. In practice, for each epoch and filter, we derived the flux at the effective wavelength. When no observation in a given filter/epoch was available, the missing measurement was obtained through interpolation of the light curve in the given filter or, if necessary, by extrapolating the missing photometry assuming a constant colour from the closest available epoch. The fluxes, corrected for extinction, provide the spectral energy distribution at each epoch, which is integrated by the trapezoidal rule, assuming zero flux at the integration boundaries. The observed flux is converted into luminosity for the adopted distance. 

We note that, shortly after discovery, the SN was not detected in radio setting a fairly restrictive upper limit to the radio luminosity \citep{Stockdale:2012} and that the  $X$-ray detection, even neglecting the probable contamination by nearby sources,
corresponds to a negligible contribution to the bolometric flux of about $1 \times 10^{39}\,{\rm erg}\,{\rm s}^{-1}$ \citep{Pooley:2012a}, that is $<1\%$ of the bolometric flux in the first month after explosion.

\begin{figure}
\includegraphics[scale=.44,angle=0]{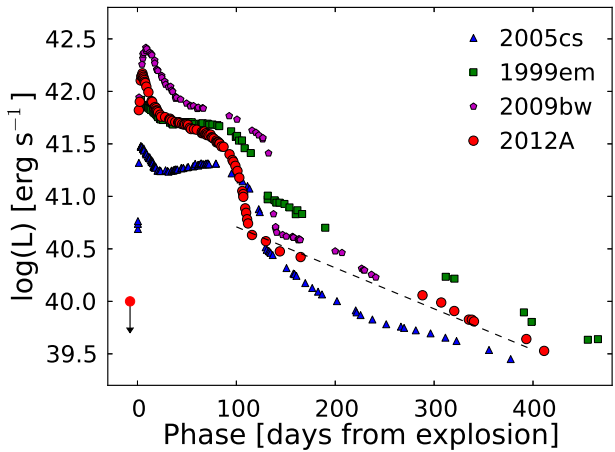}
\caption{Bolometric light curve of SN 2012A, computed integrating the fluxes from the $UV$ to near $IR$ bands, 
compared with those of the type IIP SNe 2005cs, 1999em and 2009bw. The dashed line shows the slope of the $^{56}$Co decay.} 
\label{bol}
\end{figure}

The bolometric light curve is presented in Figure~\ref{bol} together with those of SNe 1999em, 2005cs and 2009bw computed with the same prescription. SN~1999em was selected as reference because it provided the best match with the early spectrum of SN 2012A as given by the automatic classification tool (cf. Section~2), while SNe 2005cs and 2009bw were chosen because they appear to encompass the observed properties of SN 2012A  being fainter and brighter than SN~2012A, respectively. In fact it appears that the early luminosity of SN~2012A well matches that of SN~1999em.

We emphasize that at very early phases the far $UV$ spectral range contributes almost 50\% of the total bolometric luminosity, but in two weeks it drops to less than 10\%. This should be taken into account when comparing with other SNe without far $UV$ coverage. 

As noted before, in SN~2012A the {\em plateau} luminosity is not really constant showing a monotonic decline up to $80-90$ days from explosion. The drop, marking the end of the hydrogen envelope recombination, begins $15-20$ days earlier than in most type IIP SNe, including SNe~1999em and 2005cs, and the late linear tail sets in earlier.  Also we note that in SN~2012A the drop in luminosity from the plateau to the light curve tail is deeper than in SNe~1999em or 2005cs and similar to SN~2009bw. Moreover the luminosity of the linear late tail is intermediate between those of SNe 1999em and 2005cs, resembling SN~2009bw. As mentioned previously, in typical type II SNe the linear tail is powered by the energy  input from $^{56}$Co decay. This is confirmed by the observed luminosity decline rate of SN~2012A that, as seen in Figure~\ref{bol}, matches fairly well the predicted decline assuming that all the high energy photons and positrons from the radioactive decay are fully trapped and converted into $UV$-optical radiation. In this case, the tail luminosity is a direct indicator of the amount of $^{56}$Ni (the parent element in the radioactive chain) produced in the explosion. We can therefore conclude that the $^{56}$Ni mass of SN~2012A was about half that of SN~1999em and about twice that of SN~2005cs.

For a more accurate determination of the $^{56}$Ni mass it is convenient to refer to SN~1987A for which the mass of radioactive nichel has been accurately estimated to be 
$M(^{56}$Ni) = 0.075$\pm0.005\, {\rm M_{\odot}}$ \citep{danziger:1988, woosley:1989}. The comparison of the bolometric light curves of SNe 2012A and 1987A, computed integrating  the luminosity in the same spectral range,  shows that $L(2012{\rm A})/L(1987{\rm A}) = 0.14$  from which we derive $M(^{56}$Ni) = 0.011$\pm 0.004\, {\rm M_{\odot}}$ for SN~2012A, where the error budget is dominated by the uncertainty on the distance. 
 
As an independent check,  following \cite{elmhamdi:2003b}, we estimated the steepness function $S$, defined as steepness of the $V$ light curve at the inflection point during the rapid drop from the plateau to the radioactive tail. It was shown that $S$ is strongly correlated to the ejected $^{56}$Ni  mass \citep[cf. Figure~5 in ][]{elmhamdi:2003b}. From a value of $S =  0.16$~mag~d$^{-1}$ for SN~2012A, we calculate $M(^{56}$Ni) = 0.015$\, {\rm M_{\odot}}$, which is in good agreement with the value  derived from the observed luminosity in the radioactive tail.
 
The ejected mass of $^{56}$Ni derived for SN~2012A is intermediate between those of prototypical SNe~IIP, e.g. SNe 1969L and 1988A for which $M(^{56}$Ni) = 0.07$\,{\rm M_{\odot}}$ \citep{turatto:1993}, and the
value for underluminous  SNe~IIP \citep{pastorello:2004}, e.g. SNe~1997D and 2005cs, for which $M(^{56}$Ni) = 0.002\, ${\rm M_{\odot}}$ and 0.003\, ${\rm M_{\odot}}$, respectively \citep{turatto:1998, benetti:2001, pastorello:2009}. A similar $^{56}$Ni mass was found for SN~2009bw [$M(^{56}$Ni) = 0.022\, ${\rm M_{\odot}}$, cf. \cite{inserra:2012}]. 

\section{Spectroscopy}

Spectroscopic observations of SN~2012A were carried out with several telescopes commencing short after explosion, continued for well over one year, and yielding a total of 47 epochs of  medium/low resolution spectroscopy. In addition, high-resolution spectra were obtained at four epochs. The journal of the spectroscopic observations is given in Table~\ref{telescope_spec}. For each spectrum we report: the date, MJD and phase from explosion, the telescope and the instrumental configuration, the spectral range and, finally, the resolution, estimated from the FWHM of the night sky lines.

Data reduction was performed using standard {\sc iraf} tasks.  Spectral images were bias and flat-field corrected, before the SN spectrum was extracted by tracing the stellar profile along the spectral direction and subtracting the sky background along the slit direction. Wavelength calibration was accomplished by obtaining comparison lamp spectra, while for flux calibration we referred to the observations of spectrophotometric standard stars obtained, when possible, in the same nights as the SN. The flux calibration of all spectra was verified against $BVRI$ photometry and corrected, if necessary. The telluric absorption corrections were estimated from the spectra of spectrophotometric standards. Notwithstanding this, often imperfect removal can affect the profile of the SN features that overlap with the strongest atmospheric absorptions, in particular the telluric band at $7570-7750$ \AA. 

For the high-resolution spectra  we performed an additional check on the accuracy of the wavelength calibration by measuring the wavelength of night-sky lines \citep[O~I, OH, Hg~I and Na~I~D,][]{osterbrock:2000}. With this we achieved an accuracy of the velocity scale of  $0.1\,{\rm km}\,{\rm s}^{-1}$. 

\begin{table*}
\caption{Journal of spectroscopic observations. Phase is from the adopted epoch of explosion, MJD=$55933_{-3}^{+1}$ } \label{telescope_spec}
\begin{tabular}{c c c c c c}
\hline
Date	  &MJD      &Phase	& Instrumental & Range & Resolution        \\
&     &[d]         & configuration$^1$& [\AA]  & [\AA]		 \\ 
\hline
20120109 & 55936.14   & 3.1  &NOT+ALFOSC+gm4 &3400-9000 &14   \\       
20120110 & 55936.95   & 3.9  &Ekar+Echelle+gr300&3780-7400&0.34\\
20120115 & 55941.08   & 8.1  &Ekar+Echelle+gr300&4060-7785&0.38\\
20120115 & 55941.21   & 8.2  &INT+IDS+R150V  &3500-10000&13   \\
20120116 & 55942.14   & 9.1  &Ekar+Echelle+gr300&4060-7785&0.38\\
20120117 & 55943.14   &10.1  &INT+IDS+R150V  &3500-10000&13   \\
20120117 & 55943.97   &11.0  &OHP+1.93m+Carelec &3700-7300 &7  \\
20120118 & 55944.03   &11.0  &Ekar+AFOSC+gm4,gm2 &3500-9200 &24   \\
20120118 & 55944.14   &11.1  &INT+IDS+R150V  &3500-10000&13   \\
20120118 & 55944.98   &12.0  &OHP+1.93m+Carelec &3700-7300 &7  \\
20120119 & 55945.19   &12.2  &INT+IDS+R150V  &3500-10000&13   \\
20120120 & 55946.06   &13.1  &OHP+1.93m+Carelec &3700-7300 &7  \\
20120121 & 55947.00  &14.0  &OHP+1.93m+Carelec &3700-7300 &7  \\
20120121 & 55947.98   &15.0  &Ekar+AFOSC+gm4 &3500-8200 &24   \\
20120123 & 55949.98   &17.0  &Ekar+AFOSC+gm4 &3500-8200 &24   \\
20120124 & 55950.97   &18.0  &Ekar+AFOSC+gm4 &3500-8200 &24   \\
20120126 & 55952.15   &19.1  &Pennar+B\&C+300tr/mm &3400-7800 &10   \\
20120127 & 55953.03   &20.0  &Ekar+AFOSC+gm4 &3500-8200 &24   \\
20120129 & 55955.97   &22.9  &Ekar+AFOSC+gm4 &3500-8200 &24   \\
20120130 & 55956.92   &23.9  &Ekar+AFOSC+gm4 &3500-8200 &24   \\  
20120205 & 55962.94   &29.9  &Pennar+B\&C+300tr/mm &3400-7800 &10 \\
20120210 & 55968.11   &35.1  &CAHA+CAFOS+b200+r200&3400-10500&10   \\
20120213 & 55970.94   &37.9  &Pennar+B\&C+300tr/mm &3400-7800 &10   \\
20120218 & 55975.94   &42.9  &Ekar+AFOSC+gm4 &3500-8200 &24 \\
20120222 & 55980.00   &47.0  &Ekar+AFOSC+gm4 &3500-8200 &24   \\
20120225 & 55982.99   &50.0  &Ekar+AFOSC+gm4 &3500-8200 &24   \\
20120228 & 55985.92   &52.9  &Ekar+AFOSC+gm4 &3500-8200 &24   \\
20120306 & 55992.88   &59.9  &Pennar+B\&C+300tr/mm &3400-7800 &10   \\
20120309 & 55996.43   &63.4  &Ekar+Echelle+gr300&3780-7400&0.34\\
20120311 & 55998.09   &65.1  &CAHA+CAFOS+b200&3300-8850&13   \\
%20120313 & 55999.17   &66.2  &NTT+EFOSC2+gr11 & 3350-7450&14   \\
20120313 & 55999.18   &66.2  &NTT+EFOSC2+gr11+gr16 &3350-10000&12   \\
%20120315 & 56000.09   &67.1  &NTT+SOFI+GB    &9400-16400&20   \\
20120315 & 56000.11   &67.1  &NTT+SOFI+GB+GR    &9400-25000&20   \\
20120315 & 56001.88   &68.9  &Pennar+B\&C+300tr/mm &3400-7800 &10   \\
20120317 & 56003.91   &70.9  &Ekar+AFOSC+gm4 &3500-8200 &24   \\
20120325 & 56012.01   &79.0  &Pennar+B\&C+300tr/mm &3400-7800 &10   \\
20120327 & 56013.84   &80.8  &Ekar+AFOSC+gm4 &3500-8200 &24  \\
20120329 & 56015.83   &82.8  &Pennar+B\&C+300tr/mm &3400-7800 &10   \\
20120331 & 56017.90   &84.9  &Ekar+AFOSC+gm4 &3500-8200 &24  \\
20120412 & 56029.82   &96.8  &Pennar+B\&C+300tr/mm &3400-7800 &10   \\
20120422 & 56040.06   &107.1 &CAHA+CAFOS+g200      &4100-10200&13   \\
20120424 & 56041.91   &108.9 &Ekar+AFOSC+gm4 &3500-8200 &24  \\
20120427 & 56044.94   &111.9 &Ekar+AFOSC+gm4 &3500-8200 &24  \\
20120501 & 56048.89   &115.9 &CAHA+CAFOS+g200      &3800-10200&13   \\
20120515 & 56062.87   &129.9 &CAHA+CAFOS+g200      &3800-10200&13   \\
20120515 & 56062.89   &129.9 &INT+IDS+R150V        &4000-9500 &10   \\
20120608 & 56086.89   &153.9 &NOT+ALFOSC+gm4       &3400-9000 &14   \\
%20120627 & 56105.88   &172.9 &WHT+ISIS+R158R       &5700-10000&5 \\
20120627 & 56105.89   &172.9 &WHT+ISIS+R300B+R158R    &3500-10000 &5 \\
20130203 & 56326.05   &393.1 &TNG+LRS+LR-B,LR-R    &3350-10370&10 \\
\hline
\end{tabular}

$^1$ NOT = 2.56m Nordic Optical Telescope (La Palma, Spain); Ekar = Copernico 1.82m Telescope (Mt. Ekar, Asiago, Italy);  INT = 2.5m Isaac Newton Telescope (La Palma, Spain); OHP = Observatoire de Haute-Provence 1.93m Telescope (France); Pennar = Galileo 1.22m Telescope (Pennar, Asiago, Italy); CAHA = Calar Alto Observatory 2.2m Telescope (Andalucia, Spain); NTT = 3.6m ESO NTT (La Silla, Chile); WHT = 4.2m William Herschel Telescope (La Palma, Spain); TNG = 3.6m Telescopio Nazionale Galileo (La Palma, Spain).
\end{table*}

\subsection{Spectral evolution}\label{spec_evol}

\begin{figure*}
\includegraphics[scale=.8,angle=0]{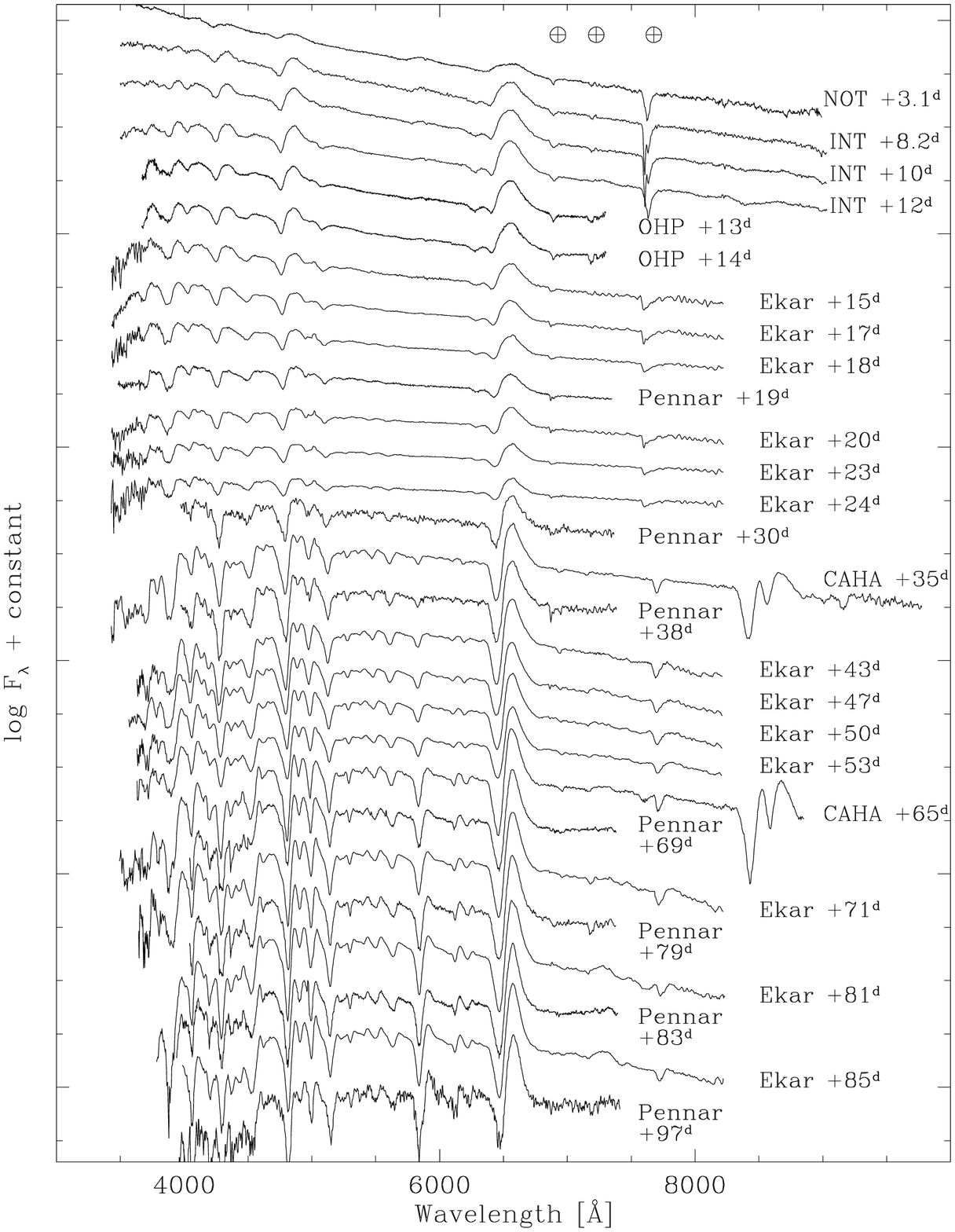}
\caption{SN 2012A: spectral evolution during the first 3 months. See Table~\ref{telescope_spec} for the reference to the telescopes labeled in the figure and the instrumental configuration. Wavelength is in the observed frame, telluric absorptions are marked with a $\oplus$ symbol. The NTT SOFI+EFOSC2 spectrum at phase +67 d (3350$-$25000 \AA\/) is shown is Figure~9.} 
\label{fig_spectra1}
\end{figure*}

\begin{figure*}
\includegraphics[scale=.8,angle=0]{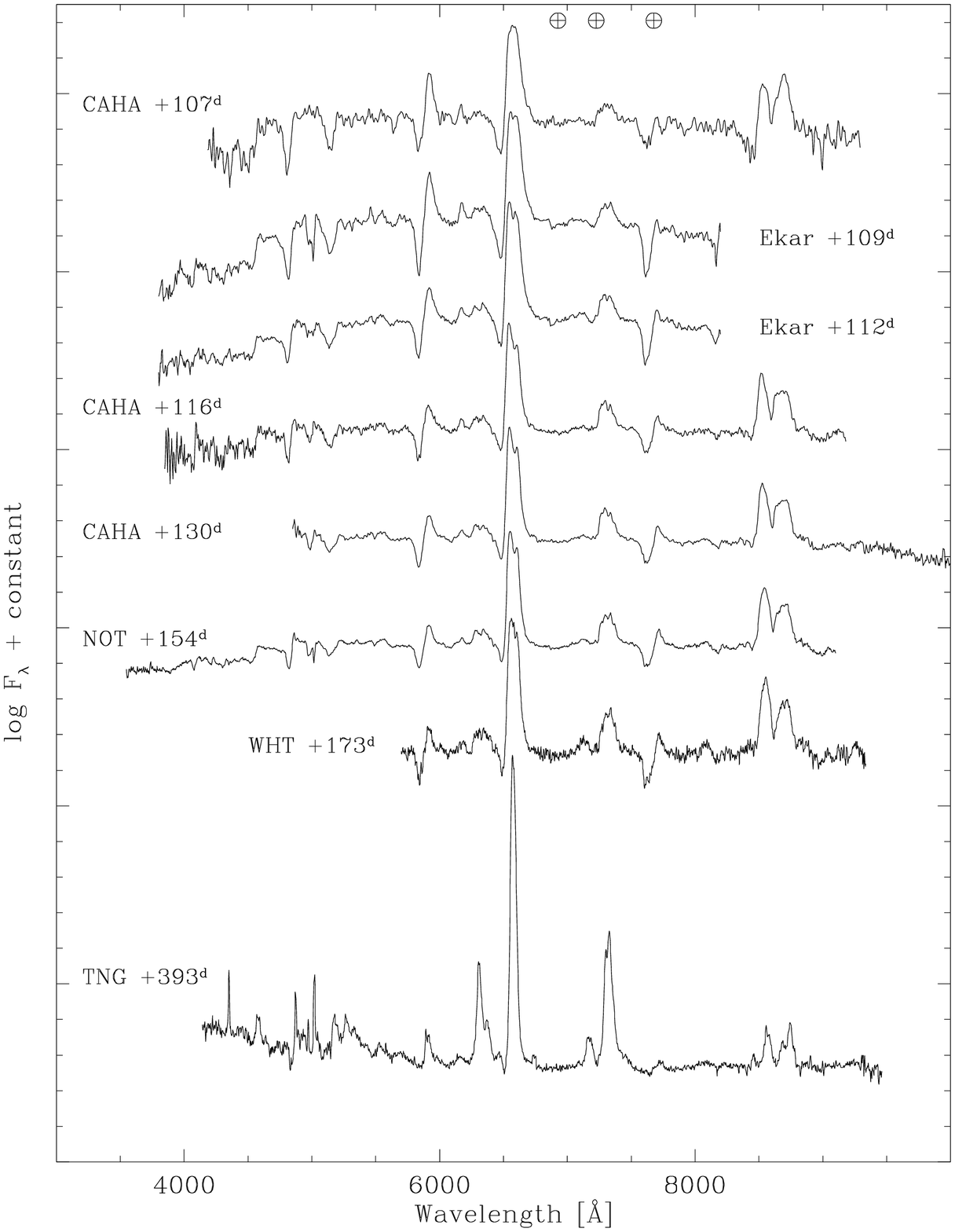}
\caption{SN 2012A: spectral evolution during the nebular phase. See Table~\ref{telescope_spec} for the reference to the telescopes labeled in the figure and the instrumental configuration. Wavelength is in the observed frame, telluric absorptions are marked with a $\oplus$ symbol.} 
\label{fig_spectra2}
\end{figure*}

The overall spectral evolution of SN~2012A is shown in Figures~\ref{fig_spectra1} and \ref{fig_spectra2}.
The first set of spectra (Figure~\ref{fig_spectra1}) illustrates  the spectral evolution during the hydrogen envelope recombination from shortly after shock breakout ($+3$ d) to the beginning of the transition phase, while the second set (Figure~\ref{fig_spectra2}) shows the transition from the photospheric to the nebular phase.
   
The earlier spectra are characterized by a very blue continuum, with blackbody temperatures above $10^4$~K. The only prominent features are the hydrogen Balmer lines with broad P-Cygni profiles, and He~I 5876~\AA. After a couple of weeks, the He~I 5876~\AA\/ disappears while the metal lines increase in strength, and after one month, become the dominant features (apart from H$\alpha$ and H$\beta$). The identified lines include  Fe~II (4500~\AA\/, 4924, 5018 and 5169~\AA\/), Sc~II (4670~\AA\/, 5031~\AA), Ba~II (6142~\AA), Ca~II (8498, 8542 and 8662~\AA, H\&K), Ti~II (blended with Ca~II H\&K).  Strong line blanketing, especially by Fe~II transitions, characterizes the blue spectral region below 3800~\AA, suppressing much of the $UV$ flux. The Na~I~D feature emerges where earlier on He~I 5876~\AA\/ was detected. 

\begin{figure}
\includegraphics[scale=0.7, angle=0]{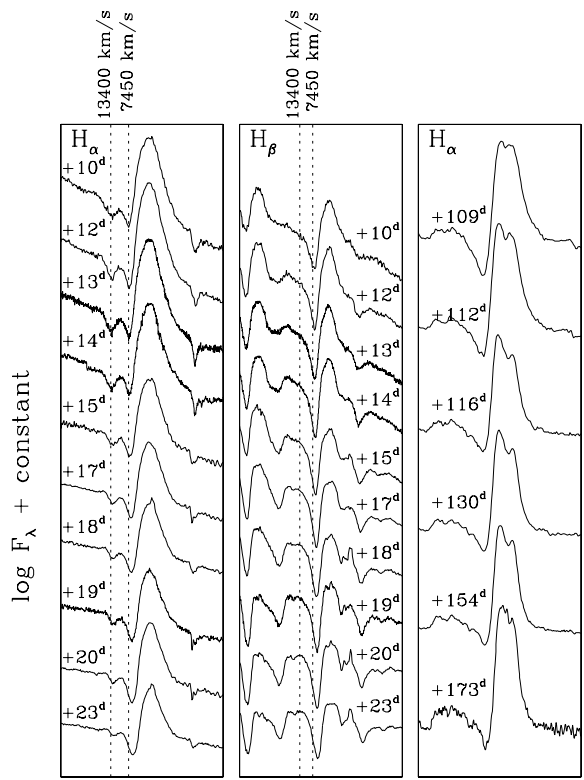} 
\caption{Evolution of the Balmer lines in SN 2012A. Left panel: evolution of H$\alpha$ during the early photospheric phase;  the two dashed lines correspond to the minimum of the P-Cygni profile of the spectrum at +10 d ($\sim$7450~km~s$^{-1}$) and to the faint absorption just to the blue of the H$\alpha$ P-Cygni absorption ($\sim$ 13400 km s$^{-1}$ if identified with H$\alpha$). Middle panel: the same as in the left panel but for H$\beta$ where there is no evidence of a similar absorption feature. Right panel: evolution of the H$\alpha$ profile during the transition from the photospheric to the nebular phase. We emphasize the asymmetry of the H$\alpha$ emission.}
\label{fig_Ha}
\end{figure}

In the spectra between +10 to +25 d, a faint  absorption feature is visible at about 6270 \AA\/, just to the blue side of the H$\alpha$ P-Cygni absorption. A similar feature was  detected by  \cite{pastorello:2006} in SN~2005cs and attributed either to high-velocity detached  hydrogen  (for SN~2012A $\sim 13800$~km~s$^{-1}$) or to the Si~II $6347-6371$~\AA\/ doublet. The lack of a similar absorption feature in the blue side of H$\beta$ (left and middle panels of Figure~\ref{fig_Ha}) supports the Si~II identification. In this case, adopting for the Si~II doublet an effective wavelength of 6355 \AA\/, the expansion velocity as measured  by fitting the P-Cygni absorption profile would be $\sim$~4400~km~s$^{-1}$ around +12 d, and decreases to 4000~km~s$^{-1}$ in the following epochs. These expansion velocities are comparable to those derived from Fe~II 5169 \AA\/ at similar phases (cf. Figure~\ref{vel_lines}).

The evolution of H$\alpha$ from +109 to +173 d is shown in the right panel of Figure~\ref{fig_Ha}.  While within 100 days from explosion H$\alpha$ was characterized by a symmetric  P-Cygni profile, by day $109-112$ a double peaked emission starts to develop and becomes more evident in the subsequent spectra at phases $116-154$ d. A complex profile is seen in H$\beta$ as well, even though the blend of several lines prevents the recognition of multiple components there. H$\alpha$ asymmetry was observed before in SN~1987A \citep{phillips:1991, chugai:1991}, SN~1999em \citep{elmhamdi:2003} and SN~2004dj \citep{chugai:2005}. For the first two SNe an asymmetric,  redshifted H$\alpha$ line was reported, while SN 2004dj showed a double peaked structure. 
A similar H$\alpha$ profile clearly arises in SN~2012A by day +112, with a dominant blue peak shifted by $\sim$ $-$1000~km~s$^{-1}$ and a red one shifted by $\sim$ $+$1300~km~s$^{-1}$. As suggested by \cite{chugai:2005} and \cite{chugai:2006} for SN 2004dj, this may indicate an asymmetric (bipolar) ejection of $^{56}$Ni in the otherwise spherically-symmetric envelope. Moreover, as for SN~2004dj, the prominence of the blue peak over the red one can be  explained with a $^{56}$Ni distribution that is skewed towards the observer. In the following spectra of SN~2012A, the H$\alpha$ profile shows some evolution, with the blue peak receding to $-$500~km~s$^{-1}$ and the red peak increasing to $\sim$ $+$1800~km~s$^{-1}$.  Finally, in the late nebular spectrum (+394 d, see Figure~\ref{fig_spectra2}), H$\alpha$ shows a single narrow Gaussian profile. We note that the [Ca~II] 7391, 7324~\AA\/ features seem to show a similar structure, although we cannot test with the [O~I] doublet (6300, 6364 \AA\/) as this line becomes prominent only in our last spectrum at phase +394 d. 

\begin{figure*}
\includegraphics[scale=.8,angle=0]{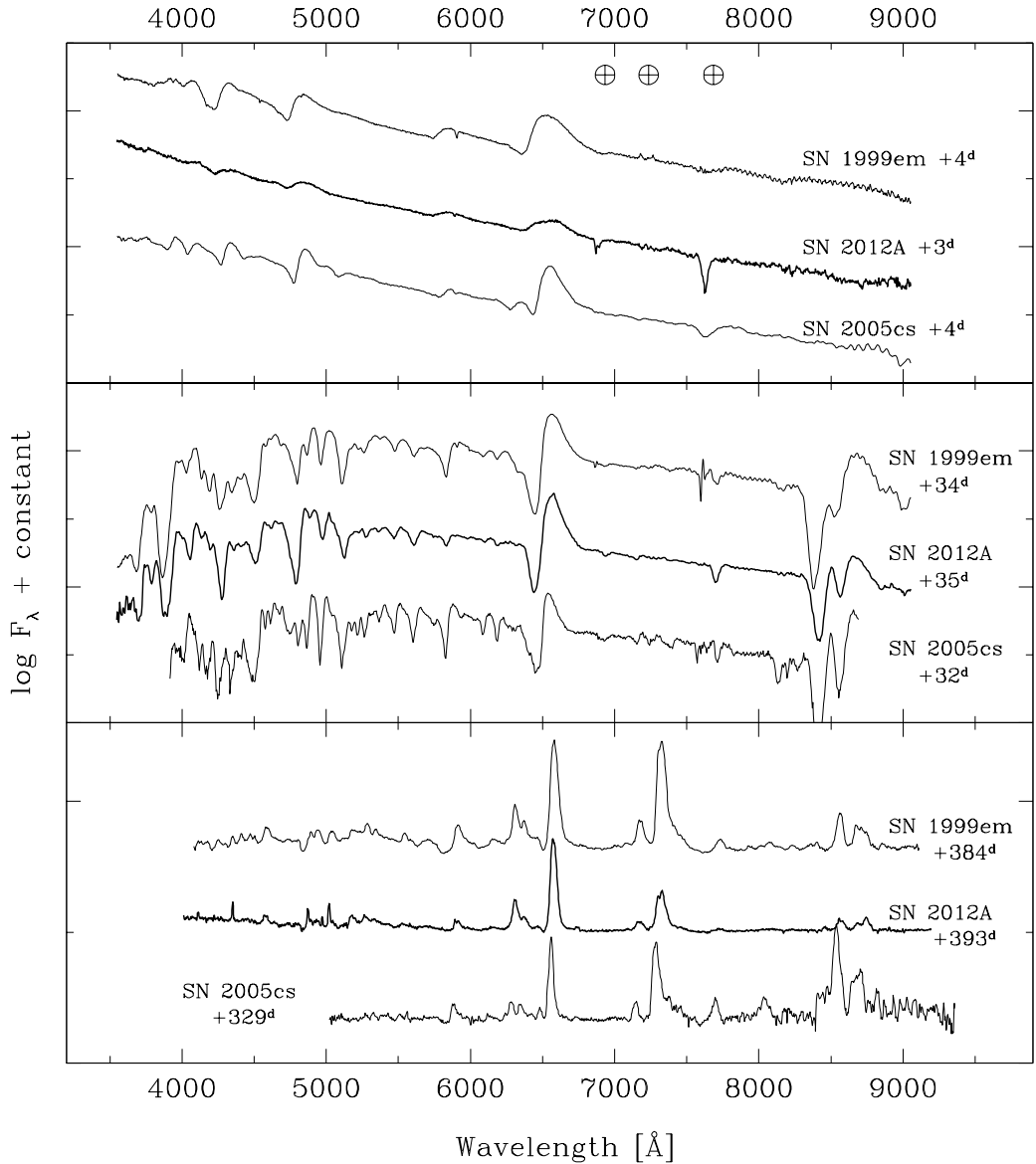}
\caption{Comparison among spectra of SN 2012A, SN 1999em and SN 2005cs at similar phases. Top panel: few days past explosion; middle panel: about one month past explosion; bottom panel: about one year past explosion.} 
\label{spec_comparison}
\end{figure*}

In Figure~\ref{spec_comparison} we compare the spectra of SN 2012A at about maximum, one month and one year after explosion with those of SN 1999em \citep{elmhamdi:2003} and SN~2005cs \citep{pastorello:2006} taken at comparable phases. Several similarities with these SNe~IIP, and in particular with SN~1999em, have already been remarked upon. All CC~SNe  are characterized  by blue continua at the early phases with blackbody temperatures that become very similar after about one month. These three SNe also show similar spectral features, with Balmer lines and He~I 5876~\AA\/ early on, and one month later the emergence of metal lines, in particular Fe~II , Sc~II, Ba~II and the Ca~II infrared triplet.

During the transition from the plateau to the radioactive tail  (Figure~\ref{fig_spectra2}), the continuum emission remains strong. This can be seen from the deep absorption of Na~I and a new strong feature of O~I at 7774 \AA. 
At the same time nebular emission lines begin to emerge, first of all [Ca~II] $7291, 7324$~\AA\/ and subsequently [O~I] $6300, 6364$~\AA\/ and [Fe~II] $7155, 7171$~\AA.
The last spectrum taken at +394~d is well in the nebular phase. Along with the always strong  H$\alpha$ line, with FWHM $\simeq 2000\, {\rm km}\,{\rm s}^{-1}$, the most prominent lines are permitted emissions from H$\beta$, Na~I~D and the Ca~II infrared triplet and forbidden lines of [Ca~II], [O~I], [Fe~II] and Mg~I] 4572~\AA. Additional emission in the range $5000-5500$~\AA\/ and around 7710~\AA\/ is most likely due to [Fe~I] and [Fe~II] multiplets. We also notice that bluewards of 4500~\AA\/, 
the spectrum shows the signature of a blue continuum that most likely is due to contamination from a nearby hot star or association.

In Table~\ref{neb} we report the measurements of the flux of the most prominent lines detected in the latest spectrum. Nebular line fluxes for a sample of SNe~IIP  have been collected by \cite{Maguire:2012} showing only a moderate variation from object to object. Actually, for SN~2012A the Ca~II and O~I emission relative to H$\alpha$ are among the weakest of the sample for the given phase, whereas the Fe~II/H$\alpha$ emission ratio is close to the average. 
The ratio of the two lines in the [O~I] 6300, 6364 doublet is $2.6$, similar to SN~2004et and  close to the expected value for the thin regime, whereas for most SNe~II at this phase the ratio is much lower signaling that the ejecta still has some optical depth.  

\cite{Jerkstrand:2012} have used their spectral synthesis code to model the nebular emission line fluxes for SN explosions of different mass progenitors, and find [O~I], Na~I~D and Mg~I] to be the most sensitive lines. By comparing their models with late time observations of the type IIP SN~2004et, they found a best match with a 15 M$_{\odot}$ progenitor model. 
From Table~\ref{neb} we derive the fractions of the line luminosities normalized to  the luminosity from the $^{56}$Co decay at +394 days (assuming $M(^{56}$Ni) = 0.011$\, {\rm M_{\odot}}$ and a distance of 9.8 Mpc, cf. Sections~\ref{distance}, \ref{bolometric}), finding values of 2.6\% for [O~I] $6300, 6364$~\AA\/, 0.70\% for Na~I~D and 0.63\% for Mg~I] 4571 \AA\/. The comparison with the expected line flux from models  \citep[see Figure~8 in ][]{Jerkstrand:2012} shows that the [O~I] and Na~I~D values are close to the 15 M$_{\odot}$ track, while the Mg~I] line is closest to the 19 M$_{\odot}$ track. As discussed by  \cite{Jerkstrand:2012}, the [O~I] lines are more reliable mass indicators than the other ones.  Therefore we conclude that the best match from the nebular spectrum analysis is for a 15 M$_{\odot}$ progenitor. However, we note that the nebular models in \cite{Jerkstrand:2012} are computed for a higher mass of ejected $^{56}$Ni ($M(^{56}$Ni) = 0.062 M$_{\odot}$ as derived for SN~2004et). A lower $^{56}$Ni mass will result in lower ionization and temperature, somewhat altering the fraction of cooling done by various emission lines.

\cite{Maguire:2012} have also shown the existence of a correlation between Ni mass and ejecta expansion velocity, measured from the FWHM of the H$\alpha$ line. By using this relation for SN~2012A we derive $M(^{56}$Ni) = 0.022$_{-0.013}^{+0.034}$ M$_{\odot}$  which, within the uncertainties is consistent with our measurement based on the late luminosity.

\begin{table}
\caption{Flux measurements in the nebular spectrum at +394 d.
The spectrum has been corrected for the adopted extinction in the direction of the SN~2012A, $E(B-V)=0.037$ mag. }
\begin{center}
\begin{tabular}{lccc}
\hline
Line  & Flux [$\times10^{-14}~{\rm erg}\,{\rm cm}^{-2}\,{\rm s}^{-1}$] \\
\hline
H$\alpha$         & 4.10  \\
$[$O I$]$  6300 \AA& 0.81 \\
$[$O I$]$  6363 \AA& 0.31 \\   
$[$Fe II$]$          & 0.27 \\
$[$Ca II$]$         & 1.90 \\
Na ID              & 0.30 \\
$[$Mg II$]$           & 0.27 \\
Ca II NIR         & 1.00 \\
\hline \\
\end{tabular}
\end{center}
\label{neb}
\end{table}%

\subsection{Near-infrared spectrum}

\begin{figure*}
\includegraphics[scale=.8,angle=0]{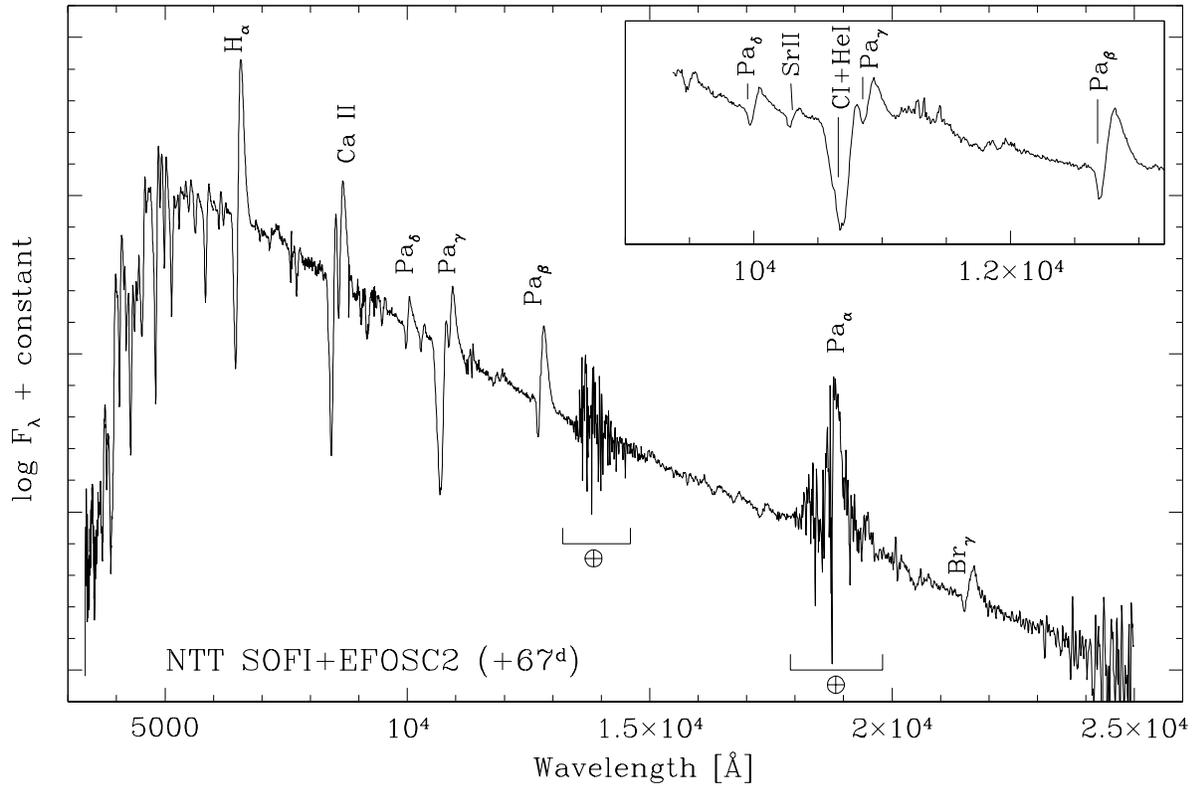}
\caption{The spectrum of SN 2012A taken with ESO NTT, EFOSC2+SOFI at phase +67 days. The insert (top right-hand corner) shows a zoomed NIR region between 9000 $-$ 13000 \AA\/ with identifications of the Paschen series, the blend of C~I 10691 \AA\/ and He~I 10830 \AA\/, and Sr~II 10327 \AA\/.} 
\label{spec_all}
\end{figure*}

A near infrared (NIR) spectrum of SN 2012A was collected with NTT+SOFI during the photospheric phase (+67 days), covering the wavelength region between 9400 to 25000 \AA. It is shown in Figure~\ref{spec_all} after merging with the +66 days optical spectrum taken with the same telescope but equipped with EFOSC2.  This was compared with the NIR SN spectra presented in \cite{gerardy:2001, fassia:2001,pozzo:2006} which were also used as guide for line identification. The NIR spectrum is dominated by the Paschen series of hydrogen, showing P-Cygni profiles similar to the Balmer lines. Br$\gamma$ is also detected. In the insert of Figure~\ref{spec_all} we show a zoomed-in-view of the NIR region  between 9000 and 13000 \AA\/. The spectral lines are typical of SNe~IIP at this phase, in particular the blend of C~I 10691 \AA\/ with He~I 10830 \AA\/ that was also observed in coeval epochs are available, i.e. SN~1997D \citep{benetti:2001}, SN~1999em \citep{hamuy:2001}, SN~2005cs \citep{pastorello:2009} and SN~2004et \citep{maguire2:2010}. Sr~II  10327 \AA\/ is clearly visible, while the contribution of Fe~II 10547~\AA\/ is not as evident as in other type II SNe \citep[see for example SN~2005cs in][]{pastorello:2009}. 

\subsection{ Blackbody temperature and expansion velocities}

\begin{figure}
\includegraphics[scale=.43,angle=0]{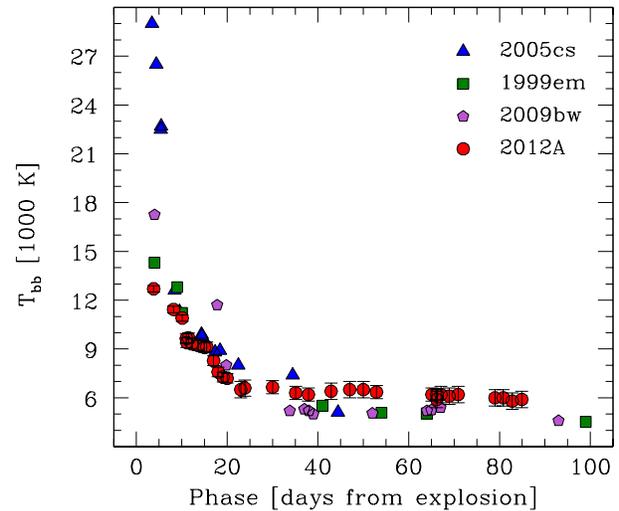}
\caption{Evolution of the continuum temperature of SN 2012A and comparison with SNe 1999em, 2005cs and 2009bw.} 
\label{tbb}
\end{figure}

Estimates of the photospheric temperatures of SN 2012A were derived from blackbody functions fitted to the spectral continuum (the spectra were corrected for the redshift and adopted extinction),
from a few days after explosion up to $\sim$80~days. Later on, due to both emerging emission lines and increased line blanketing which causes a flux deficit at the shorter wavelengths, the fitting of the continuum becomes difficult. The data are collected in Table~\ref{tbb_vel}, and the temperature evolution is shown in Figure~\ref{tbb}, along with the same measurements for  SNe 2005cs, 2009bw and 1999em for comparison.  The errors were estimated from the dispersion of measurements obtained with different choices for the spectral fitting regions. 
The early photospheric temperature of SN~2012A is above 1.2 $\times$ 10$^4$ K, but it decreases quickly to $\sim6000$ K within three weeks and then remains roughly constant. The evolution of SN~2012A is very similar to that of SNe~1999em and 2009bw, whereas SN~2005cs showed a much higher temperature at very early phases \citep[2.9 $\times$ 10$^{4}$ K,][]{pastorello:2006}. 

\begin{figure}
\includegraphics[scale=.43,angle=0]{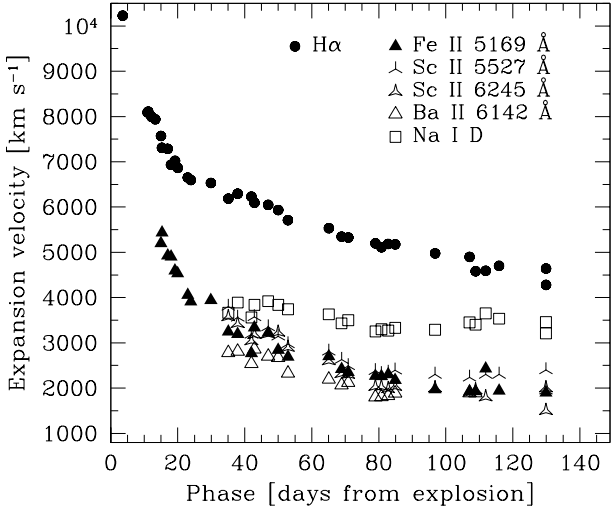}
\caption{Expansion velocity of H$\alpha$, Na~I~D, Fe~II 5169~\AA\/, Sc~II 5527, 6246~\AA\/, and Ba~II 6142~\AA\/ measured from the minima of P-Cygni profiles in the spectra of SN~2012A.} 
\label{vel_lines}
\end{figure}

\begin{figure}
\includegraphics[scale=.43,angle=0]{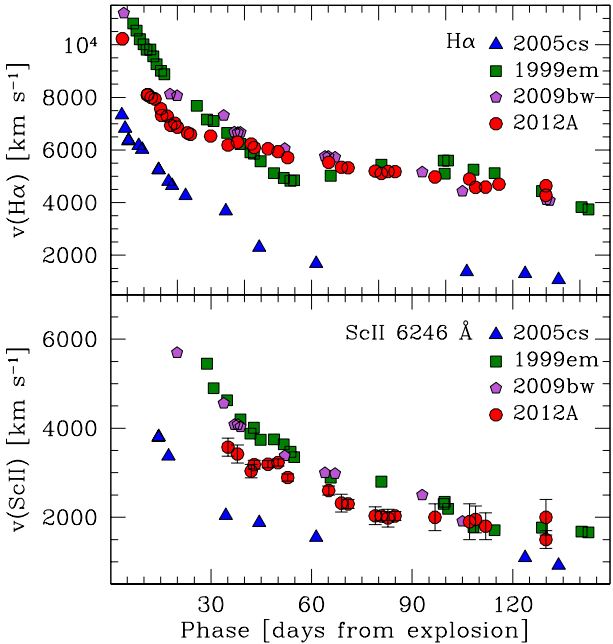}
\caption{Expansion velocities of H$\alpha$ (top panel) and Sc~II 6246~\AA\/ (bottom panel) deduced from the minima of P-Cygni profiles in the spectra of SNe~2012A, 1999em, 2005cs and 2009bw. The error bars for the velocity of H$\alpha$ in SN~2012A are smaller than the points.} 
\label{vel_ha_sc}
\end{figure}

The expansion velocity of the ejecta is measured by fitting the P-Cygni absorption components of different lines, in particular H$\alpha$, H$\beta$, Fe~II  5169 \AA\/, Sc~II 5527, 6245~\AA\/, Ba~II 6142 \AA\/ and He~I 5876 \AA\/. The latter is detected in the spectra during the first two weeks after core-collapse, while after about one month, Na~I~D emerges in the same spectral region. The expansion velocity evolution of the different lines is shown Figure~\ref{vel_lines}. H$\alpha$ displays systematically higher velocities than other lines, which is a consequence of the high optical depth of this transition. The comparison of the H$\alpha$ expansion velocity with that of other SNe~II in Figure~\ref{vel_ha_sc} shows once more that SN~2012A is more similar to SNe~1999em and 2009bw, whereas SN~2005cs has a much smaller velocity. 

If we take the velocity of the Sc~II 6245 \AA\/ line with its low optical depth as representative of the photospheric expansion velocity, we find that SN~2012A has an intermediate velocity between SNe 1999em and 2005cs (cf. Figure~\ref{vel_ha_sc}).

\begin{table*}
\caption{Measured blackbody temperatures and expansion velocities in km s$^{-1}$ (from the minima of P-Cygni absorptions) for SN~2012A. Estimated uncertainties are in parentheses.}\label{tbb_vel}
\begin{tabular}{cccccccccc}
\hline \\ 
Date	  &MJD      &Phase	& T$_{bb}$ & Fe II          & Sc II          & Sc II           &  Na ID$^1$   &  H$\alpha$ &  Ba II \\
           &               &[d]           &         [K]          & 5169 \AA\/ &  5527 \AA\/ &  6245 \AA\/ &                       &                      &  6142 \AA\/ \\
\hline
20120109 & 55936.14   & 3.1  &  12700 (200)&          &           &           &$[$8574 (200)$]$  & 10228 (200) & \\       
20120115 & 55941.21   & 8.2  &  11430 (200)&          &           &           &           &	       &\\
20120117 & 55943.14   &10.1  &  10900 (200)&          &           &           &           &	       &\\
20120118 & 55944.03   &11.0  &   9650 (300)&          &           &           &           &	       &\\
20120118 & 55944.14   &11.1  &   9400 (300)&          &           &           & $[$6013 (200)$]$&  8090 (200) & \\       
20120118 & 55944.98   &12.0  &   9680 (300)&          &           &           & $[$5896 (200)$]$&  8108 (200) & \\
20120119 & 55945.19   &12.2  &   9310 (300)&          &           &           & $[$5957 (200)$]$&  7994 (100) & \\
20120120 & 55946.06   &13.1  &   9216 (300)&          &           &           &           &  7939 (100) & \\
20120121 & 55947.00   &14.0  &   9090 (300)& 5196 (120)&          &           &           &  7569 (100) & \\
20120121 & 55947.98   &15.0  &   9136 (300)& 5432 (120)&          &           &           &  7309 (100) & \\
20120123 & 55949.98   &17.0  &   8280 (300)& 4929 (100)&          &           &           &  7286 (50)  & \\
20120124 & 55950.97   &18.0  &   7590 (300)& 4900 (80) &          &           &           & 6934 (50)   & \\
20120126 & 55952.15   &19.1  &   7270 (300)& 4597 (100)&          & &&  7021 (100)  & \\
20120127 & 55953.03   &20.0  &   7200 (300)& 4536 (50) &          & &&  6866 (80)   & \\
20120129 & 55955.97   &22.9  &   6500 (500)& 4059 (50) &          & &&  6651 (50)   & \\
20120130 & 55956.92   &23.9  &   6600 (500)& 3913 (100)&          & &&  6596 (50)  & \\
20120205 & 55962.94   &29.9  &   6650 (400)& 3943 (200)&          & &&  6532 (200) & \\
20120210 & 55968.11   &35.1  &   6300 (400)& 3247 (50) & 3804 (100)& 3575 (200)& 3664 (150) & 6185 (50)   & 2783 (200)\\
20120213 & 55970.94   &37.9  &   6200 (400)& 3189 (100)& 3563 (150)& 3421 (200)&  3890 (150)&  6295 (100) & 2813  (200)\\
20120218 & 55975.94   &42.9  &   6400 (500)& 3340 (50) & 3574 (100)& 3182 (100)& 3840 (100) & 6094 (50)   & 2861  (200)  \\
20120222 & 55980.00   &47.0  &   6500 (500)& 3224 (50) & 3351 (80) &3192 (80)   & 3920 (100) & 6048 (50)   & 2700 (200)\\
20120225 & 55982.99   &50.0  &   6500 (500)& 2841 (50) & 3146 (80)  &3229 (80)   & 3840 (100) & 5934 (50)   & 2675 (200)\\
20120228 & 55985.92   &52.9  &   6350 (400)& 2690 (50) & 2977 (80)  &2894 (80)   & 3743 (100) & 5710 (50)   & 2325 (200)\\
20120311 & 55998.09   &65.1  &   6200 (400)& 2702 (50) & 2809 (80)  &2600 (100)  & 3630 (100) & 5532 (50)   & 2193 (200)\\
20120313 & 55999.17   &66.2  &   6200 (300)& 2668 (50) & 2815 (100)&2482 (100)  & 3756 (100)  &5511 (50)   & 2072 (100)\\ 
20120313 & 55999.18   &66.2  &   5900 (500)&          &           &           &           &5560 (50)       & 2260 (200) \\
20120315 & 56000.09   &67.1  &   6200 (500)&          &           &           &           &	       &\\
20120315 & 56001.88   &68.9  &   6100 (500)& 2417 (50) & 2627 (150) & 2318 (200) & 3432 (150) & 5345 (80)   & 2066 (200)  \\
20120317 & 56003.91   &70.9  &   6200 (500)& 2354 (50) & 2489 (80)  &2300 (100)  & 3500 (100) & 5326 (50)   & 2119 (200)\\
20120325 & 56012.01   &79.0  &   6000 (500)& 2264 (100)&  2380 (150) & 2033 (200) & 3253 (150) & 5198 (80)   & 1802 (300)\\
20120327 & 56013.84   &80.8  &   6000 (500)& 2261 (50) & 2315 (80)  &2030 (100)  & 3300 (100) & 5112 (50)   & 1807 (200)\\
20120329 & 56015.83   &82.8  &   5800 (500)& 2303 (80) & 2348 (150) & 1983 (200) & 3272 (150) & 5185 (80)   & 1832 (300)\\
20120331 & 56017.90   &84.9  &   5900 (500)& 2174 (50) & 2391 (80)  &2030 (100)  & 3330 (100) & 5176 (50)   & 1885 (200)\\
20120412 & 56029.82   &96.8  &             & 1971 (150)&  2300 (250)&  2000 (300)&  3288 (200)&  4976 (200) &\\       
20120422 & 56040.06   &107.1 &             & 1930 (150)&  2239 (200)&  1900 (400)&  3453 (300)&  4897 (100) & 1890 (30)\\
20120424 & 56041.91   &108.9 &             & 1930 (100)&  1961 (100)&  1953 (300)&  3400 (200)&  4582 (150) & 1890 (30)\\
20120427 & 56044.94   &111.9 &             & 2433 (100)&  2300 (250)&  1800 (300)&  3650 (300)&  4591 (150) &\\       
20120501 & 56048.89   &115.9 &             & 1943 (120)&  2300 (250)&           &  3533 (300)&  4701 (150) & \\       
20120515 & 56062.87   &129.9 &             &           &  2400 (200)&  1500 (200)  &  3460 (300)&  4641 (150) & \\       
20120515 & 56062.89   &129.9 &             & 1901 (150)&  1950 (100)&  2000 (400) &  3212 (300)&  4280 (150) & \\       
\hline \\ 
\end{tabular}
$^1$Note: the velocity in square brackets are those of the line identified as He~I 5876 \AA\/.
\end{table*}

\subsection{Extinction}\label{extinction}

The four high resolution spectra ($R\sim17000$) taken with the REOSC Echelle Spectrograph, mounted at Copernico 1.82m Telescope, were used to search for the presence of narrow interstellar and circumstellar lines.

We identified both the Galactic and the host galaxy Na~I~D absorption components. For the Galactic component, detected at  $-20\,{\rm km}\,{\rm s}^{-1}$, we measured the EW of the D$_1$ and D$_2$ lines, finding EW(D$_1$)=0.127$\pm$0.005~\AA\/ and EW(D$_2$)=0.204$\pm$0.005~\AA.
The intensity of the Na~I~D lines is correlated with the amount of extinction.
Applying the relation derived by \cite{poznanski:2012} for both lines combined (their equation~(9)), we obtain $E(B-V)=0.034^{-0.006}_{+0.008}$ mag. This estimate is in excellent agreement with the value derived from the \cite{schlafly:2011} recalibration of the \cite{schlegel:1998}  infrared-based dust map that is $E(B-V)=0.028$ mag. 

The weak absorption features measured at a velocity of $\sim 770$ km  s$^{-1}$ correspond to the Na~I~D in the host galaxy. These features show multiple components, for a total 
EW(D$_1+$D$_2$) = $0.03\pm0.01$~\AA. Based on this analysis we can adopt a total, Galactic plus host, extinction in the direction of SN~2012A of $E(B-V)=0.037^{-0.006}_{+0.008}$ mag, or $A_V=0.12$ mag and $A_B=0.15$ mag for a standard $R_{V} = 3.1$ reddening law \citep{cardelli:1989}.

\begin{figure}
\includegraphics[scale=.45,angle=0]{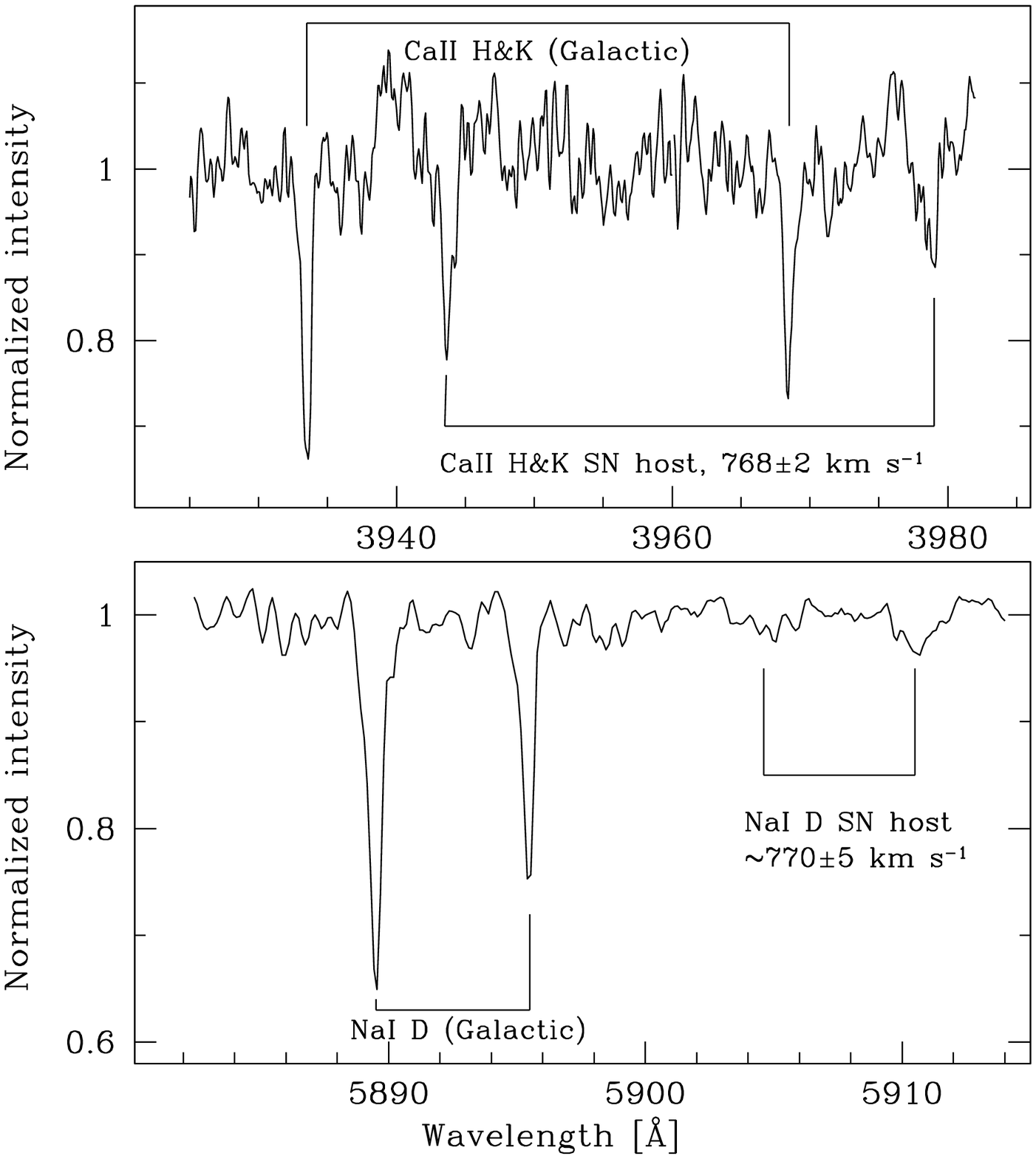}
\caption{Echelle spectrum of SN~2012A at MJD = 55936.95 (+4 d) in the Ca~II~H\&K (top panel) and Na~I~D (bottom panel) wavelength regions. Galactic Ca~II~H\&K and Na~I~D absorption features are marked.  The Ca~II~H\&K SN host absorption systems have multiple components with a mean recessional velocity
of 768$\pm$2 km  s$^{-1}$. Weak absorption features measured at a mean velocity of $\sim770$ km  s$^{-1}$ and corresponding to the Na~I doublet of the host galaxy are also detected. } 
\label{ca}
\end{figure}

In our first Echelle spectrum, the Galactic Ca~II~H\&K 
absorptions (Ca~II~H 3968.5 \AA\/ and Ca~II~K 3933.7 \AA\/) are also detected at $-20\,{\rm km}\,{\rm s}^{-1}$, along with two absorption features
that we identify with Ca~II~H\&K in the host galaxy (Figure~\ref{ca}). The Ca~II~H\&K SN host absorption systems have multiple components (at least five) with a mean recessional velocity of 765$\pm$3 km  s$^{-1}$  and 771$\pm$3~km~ s$^{-1}$ respectively. 
The association of this complex absorption systems to NGC~3239 (as well as the weak Na~I~D features at  $\sim770$~km~s$^{-1}$)
is consistent with the reported galaxy recessional velocity \citep[$753\pm3\,{\rm km}\,{\rm s}^{-1}$; ][]{devau:1991}, and that measured from the H~II
region close to the SN position ($807\pm1\,{\rm km}\,{\rm s}^{-1}$). 

\subsection{Host environment properties}

We exploited our high resolution spectra to derive the oxygen abundance of the H~II region near SN~2012A using  the relations 
 described by \cite{pettini:2004}. These authors calibrate the oxygen abundances of extragalactic H~II regions for which the oxygen abundance was determined with the direct $T_e$ method against the $N2$ and $O3N2$ indices: $N2 \equiv$ log ([N~II]6583 \AA\//H$\alpha$) 
and $O3N2 \equiv$ log$\{$ ([O~III]5007 \AA\//H$\beta$)/([N~II]6583 \AA\//H$\alpha$) $\}$. These indices have the advantage over other types of calibrations of a small wavelength separation between the lines, and hence are less sensitive to extinction and to spectral flux calibration. \cite{pettini:2004} remark that the $N2$ index yields estimates of the oxygen abundance accurate to within $\sim$0.4 dex at the 95\% confidence level while with the $O3N2$ index the oxygen abundance can be deduced to within $\sim$0.25 dex (again, at the 95\% confidence level). 

\begin{figure}
\includegraphics[scale=.45,angle=0]{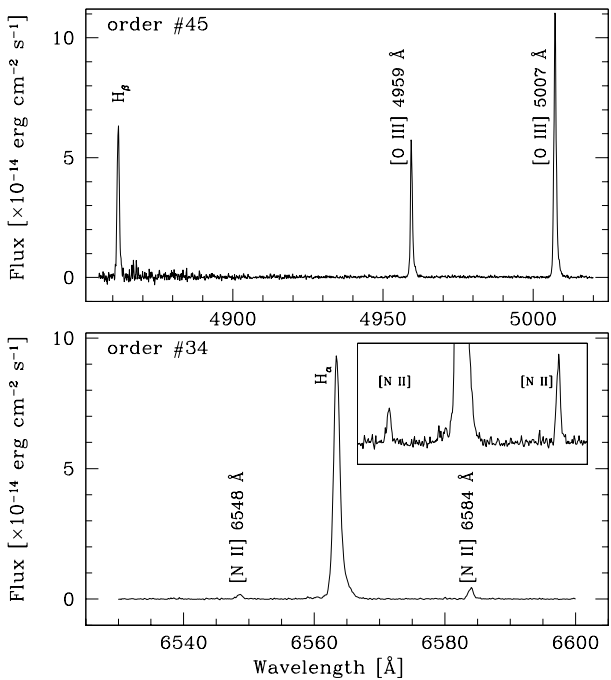}
\caption{Sections of Echelle spectrum of H~II region near SN~2012A at MJD = 55936.95 (+4 d), order \#45  (top panel), showing nebular lines H$\beta$ and [O~III] 4959, 5007~\AA\/, and order \#34 (bottom panel), with nebular lines  H$\alpha$ and [N~II] 6548, 6583~\AA\/. The insert shows a zoomed view of [N~II] 6548, 6583~\AA\/ lines.} 
\label{host}
\end{figure}

In order to obtain the spectrum of the SN host H~II region, we have oriented the slit of the spectrograph at a position angle PA $= -45^{\circ}$ and extracted the H~II region spectrum adjacent to the SN trace. We flux calibrated this spectrum (see Figure~\ref{host}) using spectrophotometric standards observed in the same night (MJD = 55936.95, at +4 d). Nebular line fluxes for [N~II] 6583 \AA\/, H$\alpha$, [O~III] 5007 \AA\/  and H$\beta$, were measured (Table~\ref{host_lines}) and we derived $N2 = -1.43\pm0.03$ and $O3N2 = 2.17\pm0.04$. From these indices we estimated the SN~2012A host oxygen abundance to be 12+log(O/H) = 8.11 $\pm$ 0.09 (using $N2$) and 
12+log(O/H) = 8.04 $\pm$ 0.01 (using $O3N2$). 

\begin{table}
\caption{Flux measurements of nebular lines in the Echelle spectrum of H~II region near SN~2012A at +4 d.}
\begin{center}
\begin{tabular}{lccc}
\hline
Line  & Flux [$\times10^{-14}~{\rm erg}\,{\rm cm}^{-2}\,{\rm s}^{-1}$] \\
\hline    
$[$N II$]$  6583 \AA& 0.41 \\
H$\alpha$                 &  11.00   \\
$[$O III$]$ 5007 \AA& 19.60 \\   
H$\beta$                   &  3.55 \\
\hline \\
\end{tabular}
\end{center}
\label{host_lines}
\end{table}%

%The oxygen abundance can be translated into an iron abundance by means of the relation derived by \cite{stoll:2012}, based on iron and oxygen abundances of the Milky Way %bulge, disk and halo stars, and using a solar oxygen abundance of 12+log(O/H)$_{\odot}$ = 8.69 \citep{asplund:2009}. The conversion is [Fe/H] = $-$11.2 + 1.25(12+log(O/H)), %implying an iron abundance for the site of SN~2012A of [Fe/H] = $-1.15\pm0.01$ using $O3N2$ or [Fe/H] = $-1.06\pm0.11$ using $N2$. 

Several authors have recently used the  \cite{pettini:2004} indices to obtain estimates of the environment metallicities for SNe~II  \citep{prantzos:2003, prieto:2008, anderson:2010,  sanders:2012, stoll:2012}, finding some evidence that the hosts of type Ib/c SNe have higher metallicity than those of type II and type Ia SNe. The environment of type II SNe show a mean oxygen abundances 12+log(O/H) = 8.58 (with a dispersion of $\sim0.75$ dex), as derived by \cite{anderson:2010} for a sample of about fifty type II SNe, or  8.65 measured for a sample of 36 type II SNe observed in the first year of the Palomar Transient Factory SN search \citep{law:2009, arcavi:2010, stoll:2012}.

Our measurement of the environment metallicity places SN~2012A in the metal-poor tail of the distribution of  oxygen abundances for the type II SNe \citep[cf. Figure~9 in][]{stoll:2012}, which includes samples by \cite{prieto:2008, anderson:2010, stoll:2012}. We may notice that \cite{prieto:2008} discuss the properties of some SNe which exploded in metal-poor environments, including SN~2006jc (a peculiar SN Ib/c with strong He~I emission lines), SN~2007I (a broad-lined SN Ic) and SN~2007bk (a 91T-like SN Ia). A number of type II SNe have also been discovered in  metal-poor galaxies \citep[see the list in ][]{prieto:2008}. However, to our knowledge, a follow-up monitoring campaign for these SNe was not conducted.

\section{Progenitor}

\subsection{Direct progenitor detection}\label{progenitor}

Since the advent of the {\it Hubble Space Telescope} ({\it HST}) and 8-m class ground-based telescopes with publicly searchable archives, the direct detection of CC SN progenitors has become possible for nearby ($\la 30$ Mpc) SNe \citep[as reviewed in ][]{Sma09}. With this in mind, we searched for pre-explosion images of NGC~3239 in which we might detect and constrain the progenitor of SN~2012A. While there are no {\it HST} images of NGC~3239, the galaxy was observed prior to the explosion of SN~2012A with the Near InfraRed Imager and Spectrometer (NIRI) on the Gemini North Telescope\footnote{Programme ID: GN-2006A-DD-2, PI: Michaud}.

NGC~3239 was observed on 2006 May 13 in $K'$ with the f/14 camera on NIRI, which has a pixel scale of 0.05 arcsec pix$^{-1}$ across a 51$\times$51arcsec$^{2}$ field of view. The adaptive optics (AO) facility ALTAIR was used when taking these observations, giving a corrected FWHM for point sources of around 0.15'', compared to the natural seeing of between 0.4 and 0.5 arcsec. Multiple short exposures were used to avoid saturating the detector on the near-infrared sky. As the target was a crowded field with extended background structure, a separate off-source field was observed in between observations of the target to allow the sky background to be subtracted. A total on-source exposure time of 600s was used, with the individual frames being reduced and co-added within the {\sc iraf gemini} package. A second set of images of NGC 3239 were obtained with the same instrument and filter on the same night, but using the f/6 camera to cover a larger field of view at the expense of a coarser pixel scale (120 $\times$120 arcsec$^{2}$ with 0.12 arcsec pix$^{-1}$). ALTAIR was not used for these data, which had a FWHM of $\sim$0.5 arcsec. As before, the images were reduced and co-added using the {\sc gemini} package, to give an on-source exposure time of 270s.

Ideally, to identify a progenitor in the pre-explosion $K'$-band image we would use either an AO image of the SN, or an image from {\it HST}. Unfortunately, we did not obtain an AO image of SN 2012A with our progenitor program, and instead we used the best natural seeing image from our follow-up campaign to astrometrically align to the pre-explosion data. The $J$-band image obtained with NOT+NOTCAM on 2012 May 6 is reasonably deep and has a FWHM of 0.5 arcsec, and is therefore suitable for our purpose.

We first aligned the NOTCAM $J$-band image directly to our pre-explosion $K'$ f/14 image. 27 sources common to both images were identified, and their positions measured. {\sc iraf geomap} was then used to derive a transformation between the two images, allowing for shifts, rotation and a change in pixel scale (i.e. the rscale fit geometry in {\sc geomap}). After rejecting two obvious outliers from the fit, the rms error on the transformation was 63 mas. We measured the SN position in the post explosion image using the average of the three different centering algorithms (centroid, gauss and filter) within {\sc iraf phot}, taking their standard deviation (9 mas) as the error.

\begin{figure}
\includegraphics[scale=0.4,angle=0]{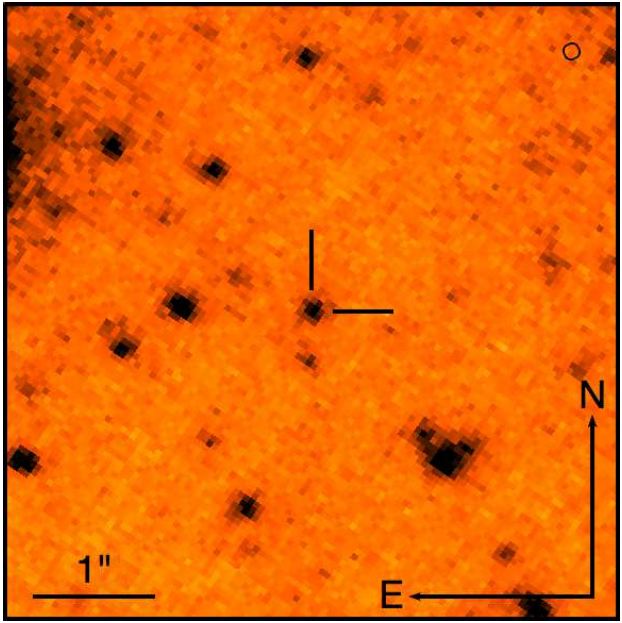}
\caption[]{A section of the pre-explosion Gemini N + NIRI + f/14  $K'$-filter image of NGC 3239, centered on the progenitor candidate. The transformed SN coordindates are at the intersection of the tick marks. Scale and orientation are as indicated in the figure; the circle in the upper right corner of the image has a radius corresponding to the 63 mas uncertainty in the SN position.}
\label{fig_progenitor}
\end{figure}

Using the pixel coordinates of the SN in the NOTCAM $J$-band image together with our geometric transformation, we determined the pixel coordinates of the SN in the pre-explosion f/14 image. A source is visible at this position, as can be seen in Figure~\ref{fig_progenitor}. As before, we measured the position of the source with three different algorithms, taking their standard deviation (1 mas) as the error. The offset between the measured pixel coordinates of the progenitor candidate and the transformed SN pixel coordinates is only 16 mas, well within our combined astrometric uncertainty of 64 mas. We hence find the progenitor candidate to be formally coincident with SN 2012A.

To determine the photometric zero point of the f/14 image, we first photometrically calibrated the f/6 image to the 2MASS system using aperture photometry of two isolated point sources in the field, for which magnitudes were listed in the 2MASS catalog. We then performed aperture photometry on five sources common to both the f/6 and f/14 images, and used these to set the zeropoint of the latter in the 2MASS system. Adding the error from both steps in quadrature, we find an uncertainty of 0.09 mag in the zeropoint of the f/14 image. We performed aperture photometry on the progenitor candidate, and using the determined zeropoint found a magnitude of $K' = 20.29 \pm0.13$ mag.  

For our adopted distance to NGC~3239 ($29.96\pm0.15$ mag) the progenitor has an absolute magnitude of $K' = -9.67 \pm 0.20$~mag. The foreground extinction ($A_V=0.115$~mag, see Section~\ref{extinction}) implies an extinction of only 0.01 mag in $K'$, with a negligible effect on the progenitor analysis. To determine a progenitor luminosity, we need to know the bolometric correction to the progenitor $K'$ magnitude. As bolometric corrections vary as a function of temperature (or spectral type), we must assume a temperature range for the progenitor. The fact that SN 2012A is a type IIP SN, with a plateau phase powered by recombination of hydrogen, implies that its progenitor must have been a hydrogen-rich red supergiant. Hence we have taken a range of bolometric corrections to the $V$-band and $V-K'$ colours based on synthetic photometry of MARCS model spectra \citep{Gus08}, appropriate for RSGs with temperatures between 3400 K and 4250 K (Fraser et al. in prep). Over this temperature range, we find that $(V-K')+BC_V$ varies between +2.89 mag for the hottest model to +2.29 mag for the coolest.

Applying the average of these values to our progenitor $K'$ magnitude, we find a bolometric magnitude of $-7.08 \pm 0.36$ mag, where the error is a combination of the photometric error, the uncertainty in the distance, and the range of plausible bolometric corrections. This corresponds to a luminosity of log $L/L_{\odot} = 4.73 \pm 0.14$~dex.

To convert the derived luminosity for the progenitor candidate to a mass, we must compare it with the predictions of stellar evolutionary models. We have made this comparison to models from the STARS code (\citealp{Eld08} and references therein), which are computed with standard prescriptions for burning, mass loss, overshooting etc. Solar metallicity models were used, although as shown by \cite{smartt:2009} the precise metallicity has negligible effect on the final luminosity of the models. \citeauthor{smartt:2009} also compare the output of the STARS code with other stellar evolutionary codes, and find good agreement.

Comparing the luminosity of the progenitor to the luminosity of stellar evolutionary models at the beginning of core neon burning, we find that a 10.5 M$_{\odot}$ progenitor is the best match. The lower limit to the luminosity is 4.6 dex, corresponding to a progenitor with a zero-age main sequence mass of 8.5 M$_{\odot}$. If the progenitor were less massive than this, it would also likely be much more luminous, having undergone second dredge-up, as discussed by \cite{Fra11}. We set a conservative upper limit to the progenitor mass of 15 M$_{\odot}$ from comparing the maximum luminosity of the progenitor to the luminosity of the STARS models at the end of core He burning.

The progenitor mass found for SN 2012A, 10.5$_{-2}^{+4.5}$ M$_{\odot}$, is comparable to the low progenitor masses found for other type IIP SNe \citep{Sma09}. We note that the fact that our progenitor detection is in $K$ makes it much less sensitive to extinction. As found for SN 2012aw by \cite{Fra12} and \cite{Van12}, and discussed in detail by \cite{Koc12}, there is evidence that some type IIP SNe have circumstellar dust which is destroyed in the SN explosion, and so the extinction towards the SN cannot be taken as a measure of the extinction towards the progenitor. However, given that the typical extinction of a few magnitudes in $V$ seen towards Galactic RSGs by \cite{Lev05} would correspond to a few {\it tenths} of a magnitude in $K$ (a value which is well within our uncertainties), this is not of great concern for this SN.

\subsection{Hydrodynamical modeling}\label{modeling}
                                                           
An independent approach to constrain the SN~2012A progenitor's physical properties 
at the explosion, namely the ejected mass, the progenitor radius and the explosion 
energy, is through the hydrodynamical modeling of the SN observables, i.e.~bolometric light curve, evolution of line velocities and continuum temperature 
at the photosphere. We adopt the same approach used for 
other CC~SNe (e.g.~SNe 2007od, 2009bw, and 2009E; see \citealt[][]{inserra:2011, inserra:2012},
and \citealt[][]{pastorello:2012}), in which a simultaneous $\chi^{2}$ fit of 
the observables against model calculations is used.

For computing the models we employ two codes: i) a semi-analytic code \citep[described in detail 
by][]{zampieri:2003} which solves the energy balance equation for an envelope with constant density in homologous 
expansion, and ii) the general-relativistic, radiation-hydrodynamics 
Lagrangian code presented in \cite{pumo:2010} and \cite{pumo:2011}. 
The latter is able to simulate the evolution of the physical properties of the ejected material 
and the behavior of the main observables up to the nebular phase, solving the equations of 
relativistic radiation hydrodynamics for a self-gravitating fluid which interacts with radiation, 
taking into account both the gravitational effects of the compact remnant and the heating effects 
linked to the decays of the radioactive isotopes synthesized during the CC~SN explosion.

The semi-analytic code is used to perform a preparatory study aimed at individuating the 
parameter space describing the CC~SN progenitor at the explosion and, consequently, to guide 
the more realistic, but time consuming simulations performed with the general-relativistic, 
radiation-hydrodynamics code.

\begin{figure}
\includegraphics[scale=.42]{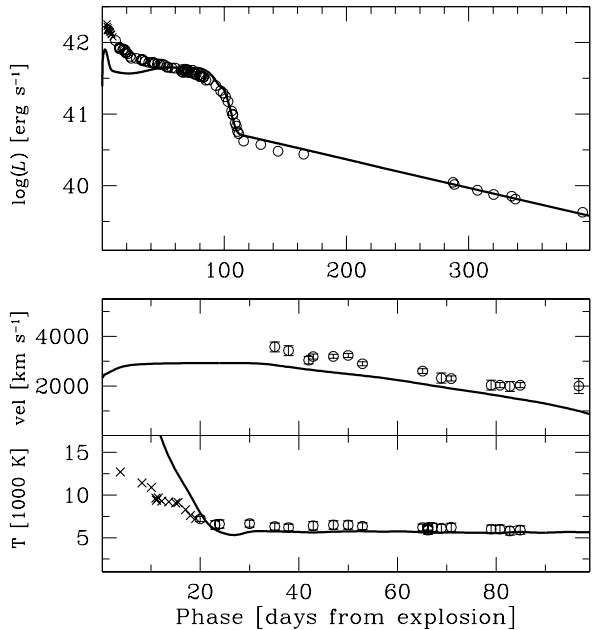}
\caption{Comparison of the evolution of the main observables of SN 2012A with the best-fitting 
model computed with a general-relativistic, radiation-hydrodynamics code (total energy $0.48$ foe, 
initial radius $1.8 \times 10^{13}$ cm, envelope mass $12.5\,{\rm M_{\odot}}$). Top, middle, and 
bottom panels show the bolometric light curve, the photospheric velocity, and the photospheric 
temperature as a function of time. To better estimate the photosphere velocity from observations, we use 
the minima of the profile of the Sc~II lines. The crosses indicate the observational data taken 
at early phases ($<$ 20 d) not considered in the $\chi^{2}$ fit (see text for details).} 
\label{model}
\end{figure}

In performing the $\chi^{2}$ fit, we do not include the observational data taken at phase $<20$ days that is approximately the time needed for the bolometric light curve to relax to the plateau. At earlier epochs all the observables are significantly affected by emission from the outermost shell of the ejecta 
\citep[cf. ][]{pumo:2011}. After shock passage, this shell, that contains only a small fraction of the envelope mass, is accelerated to very high velocities and is not in homologous expansion. The structure, evolution and emission properties of this shell cannot be probed in our simulations because at present we adopt an {\it ad hoc} initial density profile, not consistently derived from a post-explosion calculation.

In addition, we point out that our modeling is appropriate when the emission from the 
CC~SN is dominated by the expanding ejecta with no contribution from interaction with a dense circumstellar medium (CSM). In fact, as shown by \cite{moriya:2011}, the presence of a CSM around the SN (i.e. the remnant of the red supergiant wind) can heavily affect the early evolution of the light curves. In particular, following the collision of the SN ejecta with the CSM, kinetic energy  is converted to thermal energy, which is emitted as radiation. As a consequence, the SN appears $UV$ bright at early times as  was observed in SN~2009kf \citep{botticella:2010}. A similar effect may explain the bright initial $UV$ peak of SN 2012A.

The interaction of the SN ejecta with a CSM can also have signatures either in the radio or in the $X$-ray wavelength regions \citep[cf. for example][]{chakraborti:2013}. As already remarked in Section~\ref{bolometric}, SN~2012A was not detected in the radio domain \citep{Stockdale:2012}, while the $X$-ray detection corresponds to a negligible contribution to the bolometric flux \citep{Pooley:2012a}. Hence we can safely state that in SN~2012A the contribution from ejecta-CSM interaction to the observed luminosity is negligible. 

Based on the adopted explosion epoch (Section~\ref{photo_evol}), bolometric luminosity and nickel mass ($M(^{56}{\rm Ni}) = 0.011, {\rm M_{\odot}}$, Section~\ref{bol}), the best-fitting model shown in Figure~\ref{model} has a total (kinetic plus thermal)  energy of $0.48$ foe (1 foe $\equiv 10^{51}$ erg), initial radius of $1.8 \times 10^{13}$ cm, and envelope mass of $12.5\,{\rm M_{\odot}}$. This 
corresponds to a total stellar mass at the time of explosion of $\sim 14\,{\rm M_{\odot}}$, if we account for a $1.5\,{\rm M_{\odot}}$ compact remnant. The estimated uncertainty on the total stellar mass is about 15\%.

From Figure~\ref{model} (middle panel) one can notice a small ($\sim$10\%) systematic offset between the observed photospheric velocity and the modeling. This discrepancy could be reduced, at least for the first two months of evolution, by adopting for the model a slightly higher total energy of $0.5-0.6$ foe and an envelope mass between 10 and $13\,{\rm M_{\odot}}$. However in this case we would get a worse fit to the observed light curve, with a slightly shorter (by about 3 d) plateau.

\section{Discussion}

SN~2012A is an excellent SN for intensive study thanks to a number of reasons.

Our detailed photometric and spectroscopic optical monitoring, combined with near-infrared photometry and the early $UV$ photometry from {\it Swift}, allow us to reconstruct the evolution of the bolometric luminosity and derive accurate estimates of the expansion velocity and photospheric temperature. From our photometric converge during the radioactive tail of the light curve we could measure a $^{56}$Ni mass.
Our effort was facilitated by a number of favorable circumstances. Thanks to the high resolution spectra we can derive an accurate estimate of the line of sight extinction, which turned out to be very low. Also, the pre-discovery non-detections and early discovery (in the rising branch of the light curve) allow to constrain the epoch of the explosion, making SN~2012A an ideal case to test hydrodynamic models. Most importantly, thanks to fortuitous deep, high resolution, archival images and the relatively nearby distance to the host galaxy, we could directly identify a progenitor candidate for SN 2012A.

Here we focus on the attempts to constrain the progenitor mass using  the two different approaches, namely the direct detection of the progenitor in pre-explosion archival images and the hydrodynamical modeling  of the expanding ejecta.

It has been noted that, at least in some cases, the two methods lead to very different results, with the hydrodynamic modeling-derived mass being systematically higher than that obtained from direct measurements \citep{Sma09}. For instance, through modeling of SN~2005cs and SN~2004et, \cite{utrobin:2008,utrobin:2009} found progenitor masses of $\sim20$ and $\sim30\,{\rm M_{\odot}}$ respectively. These masses are a factor of $2-3$ time larger than the values obtained from an analysis of pre-explosion imaging, $\sim7$ and $\sim9\,{\rm M_{\odot}}$ respectively \citep{smartt:2009}.  In other cases the discrepancy is not so large \citep[cf. Table~7 in ][]{maguire2:2010}, but always noticeable. 

This is also the case for SN~2012A where although the two values are fully consistent within the errors, the best estimate from the hydrodynamic modeling ($\sim14\pm2\,{\rm M_{\odot}}$ including the compact remnant) is about 30\% higher that the direct mass estimate ($10.5_{-2}^{+4.5}\,{\rm  M_{\odot}}$).% In addition, it is fair to say that in both the hydrodynamic model and the direct progenitor measurements there is wide room for improvements even beyond the reported uncertainties.

\cite{utrobin:2009} speculated that hydrodynamic models could overestimate the progenitor mass due to the neglect of multi-dimensional effects. They argued that  explosion asymmetries, which are not included in the modeling, may explain the disagreement between progenitor mass determinations for SN~2004et. In that case, a possible bi-polar structure was suggested based on the line profile of  the H$\alpha$ and [O~I] 6300, 6364 \AA\/ nebular emissions \citep{chugai:2006}. A similar feature, namely a double peaked H$\alpha$ profile observed in late-stage spectra of SN~2012A (see Section~\ref{spec_evol}) could also be an evidence of the asymmetry in the SN ejecta or, most likely, in the radioactive $^{56}$Ni bubble.

For direct progenitor detections, a major source of uncertainty is the presence (or absence) of circumstellar dust. In the case of SN~2012aw, \cite{Fra12} and \cite{Van12} estimated a very low extinction after explosion, but found evidence for pre-existing circumstellar dust which was destroyed in the explosion. For SN 2012A, the possible effect of dust is reduced because the progenitor luminosity was derived from a near-infrared magnitude. A separate source of error arises from the stellar evolutionary models used. As emphasized by \cite{smartt:2009}, a different treatment of mixing processes can change the stellar core mass and hence affect the final luminosity. In particular, the choice of convective overshooting parameter will alter the final configuration of stellar models, with lower overshooting leading to lower luminosity progenitors for a given mass.
 
 As an independent constraint of the progenitor mass, we measured the relative intensities of emission lines in the nebular spectrum and compared these values with the predictions of the spectral synthesis model of
 \cite{Jerkstrand:2012}. We found that the  [OI] and Na~I~D line intensities (relative to the total radioactive energy input)  are close to the predicted values for a 15~M$_{\odot}$ star, also consistent with the estimates given above. 

In conclusion, all evidence points toward a moderate $\sim10-15$ M$_{\odot}$ red supergiant star as the progenitor of SN~2012A. 

\section{Summary}

We have presented our extensive data set for SN~2012A, starting from a few days after shock breakout, and covering over one year of evolution to when the SN had completely entered the nebular phase. The collected observations, which consist of $UBVRIJHKugriz$ photometry, low and high-resolution optical spectroscopy and a near-$IR$ spectrum, make this type IIP SN one of the most comprehensively observed ever. 

The analysis of the observations of  SN~2012A show that this is a type IIP supernova with own identity. In particular the plateau appears shorter that in  prototypical type IIP SNe, while the plateau luminosity is never really constant. The luminosity in the radioactive tail is moderate, which implies a $^{56}$Ni mass which is intermediate between those of prototypical type IIP and faint IIP SNe. In the future, a larger sample of such well studied events exploring the subtle diversity among ``normal'' SNe IIP, may help to elucidate the details of the explosion mechanism and progenitor properties.

Comparing the bolometric light  curve of SN~2012A with that of SN~1987A, we derive for SN~2012A an ejected $^{56}$Ni mass of $0.011\pm 0.004$ {\rm M$_{\odot}$}. This value is intermediate between  SN~1999em and the $^{56}$Ni-poor SN~2005cs \citep{pastorello:2009}, and close to that of SN~2009bw \citep{inserra:2012}. 

The high-resolution spectra allow us to estimate the metallicity of a H~II region close to the site of SN~2012A. The estimated oxygen abundance is 12+log(O/H) $=8.04\pm0.01$ (using $O3N2$). The derived metallicity shows that the host of SN~2012A, was in the metal-poor tail of the distribution for the hosts of type II SNe. 

A double peaked structure in H$\alpha$ profile clearly arises in SN~2012A by day +112 and disappears at around day +173. This may indicate a bipolar structure of the hydrogen excitation and can be interpreted as a result of the asymmetric ejection of $^{56}$Ni in the spherically-symmetric envelope \citep{chugai:2006}. 

The direct detection of the progenitor of SN~2012A in archival images indicates that the progenitor star had a moderate mass of 10.5$_{-2}^{+4.5}$ M$_{\odot}$. The hydrodynamical modeling of the explosion (bolometric luminosity, T$_{bb}$ and expansion velocity evolution) and the analysis of  [O~I] and Na~I~D nebular lines fluxes \citep{Jerkstrand:2012} seem to favor a slightly higher mass, around 14$-$15  M$_\odot$.  Considering the errors on the observables, the assumptions of the evolutionary models, the approximations in the hydrodynamical model, and the uncertainties introduced by a different $^{56}$Ni mass for SN 2012A compared to that in the spectral models of \cite{Jerkstrand:2012}, we find that all three mass estimates are consistent with each other.

\section*{Acknowledgments}
L.T., E.C., A.P., S.B., F.B. and M.T. are partially supported by the PRIN-INAF 2011 with the project ``Transient Universe: from ESO Large to PESSTO". 
We acknowledge the TriGrid VL project and the INAF-Astronomical Observatory of Padua for 
the use of computer facilities. M.L.P.~acknowledges the financial support from the 
PRIN-INAF 2009 ``Supernovae Variety and Nucleosynthesis Yields'' (P.I.~S.~Benetti). V.S. acknowledges 
financial support from Funda\c{c}\~{a}o para a 
Ci\^{e}ncia e a Tecnologia (FCT) under program Ci\^{e}ncia 2008 and the 
research grant PTDC/CTE-AST/112582/2009.
S.T. acknowledges support by the TRR 33 ``The Dark Universe'' of the German Research Foundation.  
G.P. and F.B. acknowledge support from ``Millennium Center for Supernova Science" (P10-064-F), with input from ``Fondo de Innovaci\'on para la Competitividad del Ministerio de Economia, Fomento y Turismo de Chile". G.P. acknowledges partial support by ``Proyecto interno UNAB DI-303-13/R". 

This paper is based on observations collected at the Copernico 1.82m Telescope and Schmidt 67/92 Telescope operated by INAF - Osservatorio Astronomico di Padova at Asiago, Italy, Galileo 1.22m Telescope operated by Department of Physics and Astronomy of the University of Padova at Asiago, Itay. Data was provided by the 2.56m Nordic Optical Telescope operated by The Nordic Optical Telescope Scientific Association (NOTSA), by 2.5 Isaac Newton Telescope and 4.3m William Herschel Telescope operated by the Isaac Newton Group of Telescopes, by the 3.6m Italian Telescopio Nazionale Galileo operated by the Fundaci\'on Galileo Galilei - INAF on the island of La Palma. The paper made also use of observations collected at the Observatoire de Haute-Provence 1.93m Telescope operated by  Universit\'e d'Aix-Marseille and CNRS, France, at the Calar Alto Observatory 2.2m Telescope operated jointly by the Max-Planck-Institut f\"ur Astronomie (MPIA) in Heidelberg, Germany, and the Instituto de Astrof\'isica de Andaluc\'ia (CSIC) in Granada/Spain, at NTT, Trappist and REM Telescopes operated by European Southern Observatory (ESO) and Prompt Telescopes operated by Cerro Tololo Inter-American Observatory (CTIO) in Chile.

\label{lastpage}
\end{document}